\documentclass[reprint,amsmath,amssymb,floatfix,aps]{revtex4-2}
\usepackage[version=3]{mhchem} 
\usepackage{graphicx}
\usepackage{enumitem}
\usepackage{xcolor}
\usepackage[normalem]{ulem}
\usepackage[fontsize=10pt]{scrextend}
\usepackage{graphicx}
\usepackage{wrapfig}
\usepackage{framed}
\usepackage{color}
\usepackage{float}
\makeatletter

\newcommand*{\rom}[1]{\expandafter\@slowromancap\romannumeral #1@}
\makeatother

\def\IB#1{\boldsymbol{#1}} 

\newcommand{\PD}[2]{\frac{\partial #1}{\partial #2}}
\def\bnabla{\boldsymbol{\nabla}}
\newcommand{\etal}{\textit{et al. }}

\begin{document}
\preprint{APS/123-QED}
\title{Pair-Interactions of Self-Propelled SiO$_{2}$-Pt Janus Colloids:\\ Chemically Mediated Encounters}
\author{Karnika Singh}
\altaffiliation{These authors contributed equally to this work.}

\author{Harishwar Raman}
\altaffiliation{These authors contributed equally to this work.}

\author{Shwetabh Tripathi}
\author{Hrithik Sharma}
\author{Akash Choudhary}
\author{Rahul Mangal}
\email{mangalr@iitk.ac.in}
\affiliation{Department of Chemical Engineering, Indian Institute of Technology Kanpur, Kanpur, India.}

\begin{abstract}
Driven by the necessity to achieve a thorough comprehension of the bottom-up fabrication process of functional materials, this experimental study investigates the pair-wise interactions or collisions between chemically active \ce{SiO2}-Pt Janus Colloids. These collisions are categorized based on the Janus colloids' orientations before and after they make physical contact. In addition to the hydrodynamic interactions, the Janus colloids are also known to affect each other's chemical field, resulting in chemophoretic interactions, which depend on the degree of surface anisotropy in reactivity and solute-surface interaction. These interactions lead to a noticeable decrease in particle speed and changes in orientation that correlate with the contact duration and yield different collision types. Our findings reveal distinct configurations of contact during collisions, whose mechanisms and likelihood is found to be dependent primarily on the chemical interactions. Such estimates of collision and their characterization in dilute suspensions shall have key impact in determining the arrangement and time scales of dynamical structures and assemblies of denser suspensions, and potentially the functional materials of the future.
\end{abstract}

\maketitle

\section{Introduction}
Artificial micro-swimmers utilize self-generated gradients to move autonomously. Due to their distinct motion characteristics that deviate from the random thermal fluctuations of Brownian motion, and their heightened responsiveness to their surroundings, 
they can serve as potential agents of drug delivery, water remediation, and microscale mixing in futuristic technologies \cite{safdar2017light, thampi2016active, luo2018micro, gao2014synthetic}. These synthetic swimmers are widely recognized for their ability to emulate key aspects of the locomotion observed in biological micro-swimmers, making them valuable models for studying biological micro-swimming.
The last couple of decades have witnessed a thrust of both numerical and analytical studies focusing on mainly two kinds of artificial swimmers. One is (i) Janus colloids: utilizing the in-built asymmetry in their physico-chemical composition, they successfully break the fore-aft symmetry to generate 
the necessary solutal gradients for their diffusiophoretic propulsion.\cite{Anderson1989} Most commonly used system is of \ce{SiO2}-Pt Janus colloids performing active motion in aqueous H$_2$O$_2$ solution due to a mechanism known as \emph{self-diffusiophoresis}.\cite{Howse2007,Ajdari2005} Other similar mechanisms such as \emph{self-thermophoresis}\cite{Palacci2015} and \emph{self-electrophoresis}\cite{Paxton2004} that use self-generated temperature and electric/ionic gradients respectively, are also well-known. The other type consists of (ii) Swimming droplets: isotropic droplets of one fluid dispersed in another immiscible fluid which utilize spontaneous asymmetry in the interfacial tension ($\gamma$) through mechanisms such as a change in surfactant activity via interfacial reactions\cite{Shashi2011,Kitahata2002,YOSHIKAWA1993} or adsorption–depletion of surfactants triggered by micellar solubilization.\cite{Izri2014,Pediredy2012, dwivedi2022self, michelin2023self} The gradient in interfacial tension results in Marangoni stress at the interface that drives the fluid from low $\gamma$ toward high $\gamma$, resulting in droplet propulsion.

Unlike random Brownian motion, individually, each class of artificial swimmers exhibits ballistic movement in a specific direction over short time intervals. However, at long time scales, transition to random motion is observed due to 
orientational thermal fluctuations.\cite{Howse2007,thutupalli2018flow,Izri2014} 
This long-time transition to random motion remains a major challenge in the successful implementation of these artificial micro-swimmers in various applications. 
Therefore, in the recent past, to gain better insights about their motion and to incorporate directionality, several studies have tried to understand the influence of various factors such as external flow,\cite{Ren2017,Sharan2022,Bishwa2020,Prateek2021_2,Palacci2015,Katuri2018} presence of chemical solutes,\cite{Prateek2021,Zhou2021, Brown2014, Gomez2016, Saad2019, Raman2023PEO, Dwivedi2023} the impact of fixed topographical features,\cite{crowdy2013, mozaffari2016self, Das2015, liu2016bimetallic} presence of tracers,\cite{Karnika2022, Katuri2021, Theeyancheri2022} or tuning the viscosity of the surrounding medium.\cite{ginot2022barrier}

These artificial micro-swimmers have also been shown to form unique dynamic self-assemblies that can naturally transform into new phases or structures.\cite{Palacci2013, ginot2018aggregation, thutupalli2018flow, hokmabad2022chemotactic, dominguez2016collective, kralchevsky2000capillary, botto2012capillary, Lippera2021, moerman2017solute, PhysRevLett.110.238301}. Together they can work to accomplish tasks that a single swimmer may not be capable of achieving in isolation \cite{maggi2016self}. 
Study of their collective behavior is also expected to provide useful insights into biological flocking/swarming,\cite{ibele2009schooling, aranson2013collective, thutupalli2018flow} predator-prey interactions\cite{meredith2020predator}, formation of bacterial colonies and multicellular bodily responses, \cite{devreotes1989dictyostelium} and inter-cellular communications \cite{goldbeter2006oscillations}. 
The collective functionality of the swimmers stems from the inter-swimmer(s) interactions. In biological swimmers, besides chemical signaling, the interactions are mostly dominated by the flow-field of the individual swimmers,\cite{pooley2007hydrodynamic} captured by the well-known squirmer model, \cite{blake1971spherical, lighthill1952squirming, zottl2014hydrodynamics} coupled with steric alignments.\cite{sokolov2012physical, zhang2010collective, schoeller2018flagellar} In artificial micro-swimmers too, hydrodynamic interactions govern most of the interactions, leading to several interesting phenomena, including viscosity reduction in suspensions,\cite{hatwalne2004rheology} synchronized motion, \cite{dombrowski2004self} and pattern formation.\cite{thutupalli2018flow}

The collective functionality of the chemically powered micro-swimmers can also be tuned by comprehending the interactions of underlying constituents, which is the subject of studies performed in the past and have been highlighted in Table \ref{table1}. In one of the first studies, Palacci \textit{et al.} performed an equilibrium characterization via the classic `Jean-Perrin' sedimentation experiment \cite{palacci2010sedimentation}. In addition to the activity-Temperature analogy, through their subsequent experiments on \ce{Pt-Au} colloids, Theurkauff \textit{et al.} found an onset of chemotactic aggregation and dynamic clusters at semi-dilute concentrations; their Keller-Segel-type model found that chemical interactions governed the clustering \cite{theurkauff2012dynamic}. 
Later,  Ginot \textit{et al.} provided a thermodynamic description of the cluster formation and reported that their evolution and kinematics were dictated by an effective adhesion energy spawning predominantly from chemical interactions \cite{ginot2015nonequilibrium, ginot2018aggregation}. 
Using a Langevin description, Saha \textit{et al.} built a coarse-grained framework for studying chemotactic aggregation in diffusion- \& reaction-limited regimes \cite{saha2014clusters}. The former condition exhibits clumping patterns and fluctuations, whereas the latter yields long-ranged instabilities. In parallel, Pöhl \& Stark performed 2D Brownian dynamics simulations of dry active suspensions interacting chemically, that influence their translational and rotational mobilities \cite{pohl2014dynamic, pohl2015self}. They showed that only when the modification to these two mobilities counteract (particles drifting towards a chemical sink, but propulsion axis turned away), a chemotactic collapse can be avoided and the dynamic clustering can be observed.

 	\begin{table*}[t!]
  \small
 		\renewcommand{\arraystretch}{0.7}
 		\begin{center}
 		\def~{\hphantom{0}}
 		\begin{tabular}{c c c c}  \\ \hline \\
        
 		Investigation & Description  & Conditions/Regime  & Result/Insight \\[10pt] \hline \\

			\begin{tabular}[t]{l} Palacci \textit{et al.} (2010) \cite{palacci2010sedimentation};\\Theurkauff \textit{et al.}\\(2012) \cite{theurkauff2012dynamic} \end{tabular}       & \begin{tabular}[t]{l}`Jean-Perrin' type expe-\\riments on settling active\\colloids\end{tabular}      & \begin{tabular}[t]{l}  \ce{Pt-Latex}; \ce{Pt-Au}\\ Upto 50\% surface\\fraction \end{tabular}     & \begin{tabular}[t]{l} $T_{\text{eff}} \sim Pe^2 $; Dynamic clustering\\above 3\% fraction, grows $\propto V$,\\governed by chemical interactions\end{tabular} \\ \\
   
            \begin{tabular}[t]{l} Ginot \textit{et al.} (2015;\\2018) \cite{ginot2015nonequilibrium, ginot2018aggregation} \end{tabular}       & \begin{tabular}[t]{l} Formulated equation of state;\\Experiments on cluster\\evolution and kinematics \end{tabular}      & \begin{tabular}[t]{l}  \ce{Pt-Au}, Up to 80\%\\area fraction; Up to\\10\% area fraction\end{tabular}     & \begin{tabular}[t]{l} Effective inter-particle `adhesion'\\$\sim$ $Pe$; $\,$ Derived aggregation-\\fragmentation rates\end{tabular} \\ \\
            
            \begin{tabular}[t]{l} Sharifi-Mood \textit{et al.}\\(2016) \cite{sharifi2016pair} \end{tabular}       & \begin{tabular}[t]{l} Semi-analytical study on\\chemo-hydrodynamic\\pair interactions\end{tabular}      & \begin{tabular}[t]{l} $Pe_{f} \ll 1$,\\non-Brownian,\\uniform mobility\end{tabular}     & \begin{tabular}[t]{l} Incidence angle and catalyst\\coverage determine assembly\\or escape\end{tabular} \\ \\
            
            \begin{tabular}[t]{l} Saha \textit{et al.} (2019) \cite{saha2019pairing} \end{tabular}       & \begin{tabular}[t]{l}Far-field framework for\\a directory of non-recip-\\rocal pair interactions \end{tabular}      & \begin{tabular}[t]{l} Quasi 2D dynamics,\\$Pe_{f} \ll 1, \; d_{cc} \gg a $ \end{tabular}     & \begin{tabular}[t]{l} Chemical interactions govern \\pairing, waltzing \& scattering \end{tabular} \\ \\
            
            \begin{tabular}[t]{l} Varma \& Michelin\\(2019)\cite{varma2019modeling}; Rojas-Pérez\\\textit{et al.} (2021) \cite{rojas2021hydrochemical} \end{tabular}       & \begin{tabular}[t]{l} Generalized \& efficient\\frameworks for semi-\\dilute suspensions\end{tabular}      & \begin{tabular}[t]{l} $Pe_{f} \ll 1$,\\non-Brownian\end{tabular}     & \begin{tabular}[t]{l} Chemo-hydrodynamic fields exhi-\\bit intricate coupling, spawning\\novel multi-body interactions  \end{tabular} \\ \\
            
            \begin{tabular}[t]{l} Sharan \textit{et al.} (2023) \cite{sharan2023pair} \end{tabular}       & \begin{tabular}[t]{l} Experiments \& model of\\surface-bound pair collision\end{tabular}      & \begin{tabular}[t]{l}  Pt$-$ \& \ce{Cu-SiO2} on\\1D crack \& 2D plane\end{tabular}     & \begin{tabular}[t]{l} \ce{Pt-SiO2} interactions are steric,\\\ce{Cu-SiO2} repel at $d_{cc} \gtrsim 7a$ \end{tabular} \\ \\
            
            \begin{tabular}[t]{c} This work \end{tabular}       & \begin{tabular}[t]{l} Pair collision experiments\\to characterize configura-\\tions and contact times\end{tabular}      & \begin{tabular}[t]{c} \ce{Pt-SiO2} on 2D plane \end{tabular}     & \begin{tabular}[t]{l} Collisions are non-steric \& occur\\over few seconds dictated by\\chemical field over $ d_{cc}<3a  $ \end{tabular}\\ \hline
         
            \end{tabular}
            
            \caption{Past studies on suspensions of Janus colloids addressing chemo-hydrodynamic interactions and their impact on emergent behavior. Here, $d_{cc}$ is the center-to-center distance between particles of radius $a$, $Pe$ is the Péclet number that denotes the persistence length of the active Brownian particle, $Pe_f$ is the Péclet number associated with solute advection via fluid flow, and $V$ is the propulsion velocity.}
 		\label{table1}
 		\end{center}
 	\end{table*}

 \normalsize

In addition to the aforementioned experiments and simulations, a series of theoretical studies in the continuum framework revealed a complex interplay of chemo-hydrodynamic interactions: a set of chemical and hydrodynamic signals, generally simplified and decoupled in the low fluid advection limit.
These fields get disturbed whenever a swimming colloid is in the vicinity of another colloid's signals, which breaks the symmetry of viscous stresses and invokes an additional kinematic response. In their semi-analytical study on pair interactions, Sharifi-Mood \textit{et al.} showed that 
assuming uniform mobility on the Janus colloid's surface, depending on the incidence angle and catalyst coverage, pair-wise interactions between solute emitting Janus colloids can result in their pairing or escape \cite{sharifi2016pair}. This assumption was relaxed by the more recent works by Michelin and co-workers\cite{varma2019modeling,rojas2021hydrochemical,traverso2020hydrochemical}, and provided two generalized frameworks for semi-dilute suspensions: one based on the method of reflections \cite{varma2019modeling} and another on force-coupling method based on multipole expansion of chemo-hydrodynamic signals \cite{rojas2021hydrochemical}. A parallel theoretical study by Saha \textit{et al.} \cite{saha2019pairing} constructed a directory of possible non-reciprocal pair interactions by exploring the parameter space of modifications to translational and rotational motion in a quasi-2D space. Dominated by chemical interactions, a dimer state (mutually chemotactic) and four dynamical states emerged from combinations of chemotactic-anti chemotactic interactions: attractively scattered, repulsively scattered, chasing, and orbits.

Despite this recent progress elucidating interactions among \ce{H2O2} fueled active Janus Particles (JPs), detailed experimental investigations on the isolated pair-wise interactions have been missing. In the decades-old context of inertially sedimenting particles\cite{fortes1987nonlinear}, meticulous investigation on the approach and departure of pair-collisions has facilitated recent insights into large-scale particle-laden turbulent flows\footnote{
Sedimenting pairs experience `draft-kiss-tumble' transition on approach, contact and departure, which is responsible for settling speeds twice as fast as mean values, high intermittency, and non-Gaussian characteristics \cite{fornari2016sedimentation}
.}\cite{brandt2022particle}.

In the same spirit, here we undertake an experimental examination, to carefully explore the approach, contact and departure during the pair-wise interactions of \ce{SiO2}-Pt JPs constrained to move in a 2D ($x$-$y$) plane. The concentration of the JPs is kept very low to avoid clustering and cross-interactions among different pairs. 
Very recently, Sharan \etal studied pair-interactions of \ce{SiO2}-Cu and \ce{SiO2}-Pt  JPs on a 1-D crack and 2-D plane. For \ce{SiO2}-Cu JPs the study demonstrated far-field repulsive interactions preventing contact. For \ce{SiO2}-Pt JPs, they reported that the interactions are mostly steric in nature and no far-field hydrodynamic or chemical interactions were observed \cite{sharan2023pair}. They also reported (\textit{i}) no significant speed fluctuations during the interaction, (\textit{ii}) 3-4 seconds of contact where JPs slide over each other, (\textit{iii}) post-collision departure due to free in-plane thermal reorientations.
In this study of \ce{SiO2}-Pt JPs on a 2D plane, while we confirm the absence of far-field chemo-hydrodynamic interactions, we observe that the collision dynamics is not simply dictated by steric interactions, instead, the near-contact chemical interactions play a major role in determining both contact time and the manner in which they slide over each other. 
With detailed characterization, we demonstrate that depending on the approach orientation of the JPs, these interactions are capable of generating significant speed and orientation fluctuations. In the end, the experimental observations regarding the rotational motion of the JPs during the collision are supported by simple qualitative estimates of the chemical torque for various pair orientations. 

\section{Materials and Methods}
\ce{SiO2}-Pt Janus particles (JPs) were synthesized using the drop-casting method as reported by Love \etal \cite{love2002fabrication} Briefly, a stock solution of 5 $\pm$ 0.35 {$\mu$m} \ce{SiO2} particles (Sigma-Aldrich, 5 wt.$\%$ solids) was diluted with de-ionized (DI) water (1:3 v/v) to prepare a 1 wt.$\%$ particle suspension. The suspension was drop-cast on a plasma-treated glass substrate. Plasma treatment was done using a plasma cleaner (PDC-32-G2, Harrick Plasma) in the presence of oxygen to make the glass slide hydrophilic, which assists in the uniform spreading of the particle suspension. A particle monolayer is formed on slow water evaporation, which was confirmed \emph{via} optical microscopy. A thin layer of platinum (Pt; thickness $\sim$ 15 nm) was then deposited on the particles via DC magnetron sputtering using a bench-top sputter coater (BT150, Hind High Vacuum Co., HHV). Due to the self-shadowing effect, only the top half of the particles get coated, making them Janus. These JPs were then dispersed in DI water using sonication. To propel the JPs, an appropriate amount of 30 wt.\% \ce{H2O2} aqueous solution (Thermo Fisher Scientific) is added to the particle mixture to achieve an overall \ce{H2O2} concentration of $\sim$3 wt.\%. For visualization an optical cell is prepared by sticking a polydimethylsiloxane (PDMS) ring of height 2 mm and diameter 1 cm on a \ce{O2}-plasma-treated glass slide. The cell is filled with the particle solution (with \ce{H2O2}) and sealed with a cover slip. Due to the higher density of JPs, they sediment towards the bottom of the cell and propel in the 2D ($x$-$y$) plane. Particles are imaged in the bright-field mode using an inverted microscope (IX73, Olympus) coupled with a CMOS camera (Oryx 10GigE, Teledyne FLIR) of resolution 1800 $\times$ 1026 pixel$^2$. A thermal stage is also mounted on the microscope to maintain a constant temperature (25 \textdegree{C}) throughout the experiment. Movies are recorded for around 5 min at 20 frames per second. After the acquisition of videos, particle tracking was done using the MOSAIC plugin in Image-J, which uses an image correlation-based approach to obtain an individual particle’s center of mass position [$x(t)$,$y(t)$] in the laboratory frame. An in-house written Python code was used to detect the orientation of JPs.

\section{Results and discussion}
\subsection{Isolated active JP motion}
\begin{figure*}[!t]
\centering
\includegraphics[width=\textwidth]{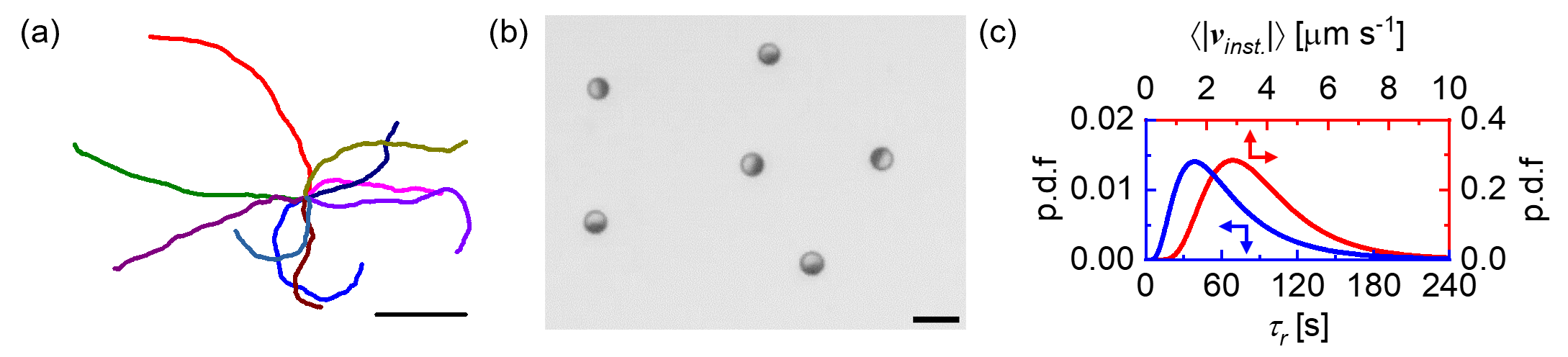}
\caption{(a) Representative trajectories ($\sim$ 60 s) of \ce{SiO2}-Pt JPs in 3 wt. $\%$ \ce{H2O2} solution. The scale bar corresponds to a length of 100 $\mu$m. (b) Bright-field optical micrograph of the active JPs. The scale bar corresponds to a length of 10 $\mu$m. (c) Probability distribution of average instantaneous speed $\langle|\textbf{v}_{inst.}|\rangle$ and reorientation timescale $\tau_{r}$ of the active JPs. Note that the distribution obtained using nearly 40 active JPs is fitted with a suitable log-normal distribution curve for better visualization.}
\label{fig:control}
\end{figure*}

Firstly, we investigate the motion of isolated 5 $\mu$m JPs in 3 wt. $\%$ \ce{H2O2} aqueous solution. Decomposition of \ce{H2O2} on the Pt patch results in a chemical gradient across the JP, causing its self-diffusiophoretic propulsion (see supporting movie S1). As seen in figure \ref{fig:control} (a), the isotropic nature of the trajectories with no preference for either $x$ or $y$ direction confirms the absence of any convective drift in JPs. 
During their active motion, the JPs maintain the half-moon orientation with their normal to the Janus plane (\textbf{n}) oriented parallel to the bottom wall, as seen in the optical micrographs shown in \ref{fig:control} (b). Hydrodynamic and chemophoretic interactions with the bottom wall are well-known to force the active JPs to adopt this stable orientation.\cite{Bishwa2020, Das2015,simmchen2016topographical} 

The rotational time scale $\tau_\text{r}$ of JPs is obtained by fitting the velocity auto-correlation function 
\begin{math}
        C(\Delta t) (= \langle\mathbf{v}_\text{inst.}(\Delta t).\mathbf{v}_\text{inst.}(0)\rangle)
\end{math} 
with the form
\begin{math}
        C(\Delta t)= 4D \delta (\Delta t)+\langle|\textbf{v}_\text{inst.}|\rangle^2 \cos{(\omega \Delta t)}\exp{\frac{-\Delta t}{\tau_\text{r}}}
        \end{math},
where $\delta$ is the Dirac delta function, and $\omega$ is the angular speed of the active JPs, if any (see supporting figure S1 (d)). The instantaneous velocity $\mathbf{{v}_\emph{{inst.}}}$ is defined as
\begin{math}
        \mathbf{v}_\text{inst.}= \frac{\mathbf{r}_\text{i+1}-\mathbf{r}_\text{i}}{t_\text{i+1}-t_\text{i}}
\end{math}. Here $\mathbf{r}_\text{i}$ is the instantaneous position vector at time $t_\text{i}$. Figure \ref{fig:control} (c) shows the Probablity distribution of $\langle|\textbf{v}_\text{inst.}|\rangle$ and $\tau_\text{r}$. The distribution is attributed to the variation in particle coating, interactions with the substrate, size variation, etc.

\subsection{Observing pair-interactions: approach and departure}
\begin{figure*}[!t]
\centering
\includegraphics[width=0.8\textwidth]{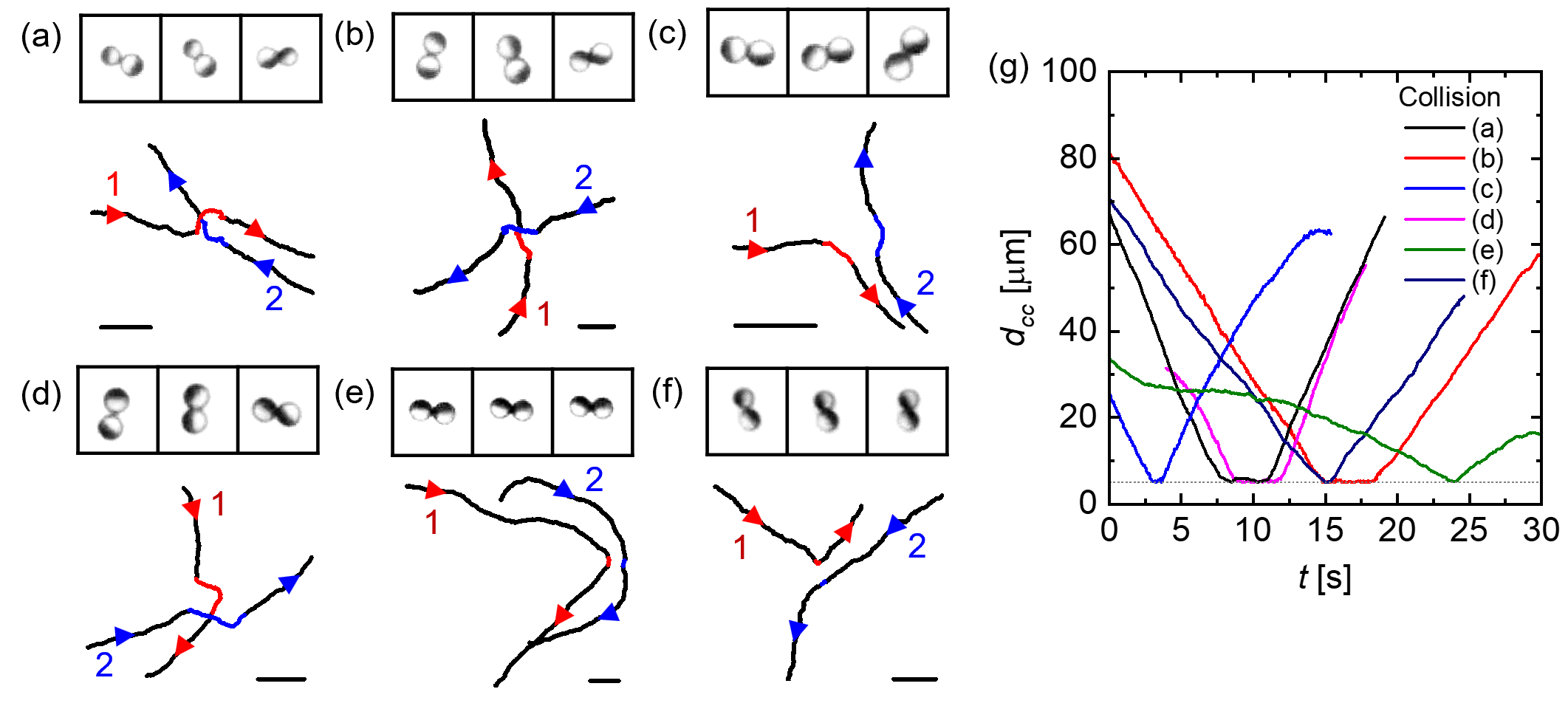}
\caption{(a-f) Representative trajectories of active JP pairs involved in different collisions. Colored segments in the trajectories indicate that the active JPs remain in contact. Time-lapse images in the insets, from left to right, show the orientation of the JPs during the first instance of contact, mid-point of collision, and just before separation, respectively. Scale bars correspond to a length of 10 {$\mu$m}. (g) Time variation of the inter-particle distance $d_\text{cc}$ for the collisions shown in figures (a-f). The dotted line indicates the size of an individual particle \emph{i.e.} the least possible value of $d_\text{cc}$.}
\label{fig:collisions}
\end{figure*}

In this study, we have carefully documented 120 pair collisions. The concentration of the particle suspension was adjusted adequately to maintain the low number density of JPs, which mostly limited the interactions among the JPs to two-body (pair) collisions only. Figures \ref{fig:collisions}(a-f) show a few representative collisions and figure \ref{fig:collisions}(g) represents the inter-particle center-to-center distance $d_\text{cc}$ as a function of time $t$ for these trajectories. From the figure, it is evident that the collisions occur in three sequential events. At first, the JPs move towards each other as $d_\text{cc}$ decreases with time $t$. Afterward, $d_\text{cc}$ reaches its minimum value ($\sim$ 5 $\mu$m) and remains constant for some time $t_\text{contact}$, during which the active JPs maintain close contact and interact by sliding or rolling against each other. These interactions occur in different forms and determine the contact length and time, which is a subject of exploration in this article.
Eventually, the JPs detach from each other and drift apart, also depicted by the subsequent increase in $d_\text{cc}$ with $t$, as shown in figure \ref{fig:collisions}(g). The optical micrographs of the colliding JPs are shown in the inset images of the figure \ref{fig:collisions}(a-f) for three distinct periods during their physical contact: the first moment of contact, the midpoint of the collision, and just before their separation. Careful inspection of the observed collisions reveals that the active JPs predominantly collide with their \ce{SiO2} hemispheres coming in contact first. Very rarely, it was observed that the Pt side of either JP comes in contact with either side of the other JP. In such a case, the approaching particles scatter instantly, as depicted in collision (f). We also observed that, towards the end of their interaction, the equators of the JPs align bringing their Pt sides closer, which triggers their separation.

The observations of approach and departure can be understood by the chemophoretic interactions originating from the \ce{H2O2} decomposition occurring on the Pt hemisphere of the active JP. 
These interactions occur due to the chemical fields generated by the neighboring active JPs in the system and usually decay as  ($\sim$
\begin{math}
        \frac{1}{{d_\text{{cc}}^{2}}}
\end{math}).\cite{golestanian2007designing,sharifi2016pair,liebchen2021interactions} Whenever the reactive side of the JP approaches another particle's silica side, the accumulation of the reaction products in the interstitial region forces the JPs to move away from each other. This repulsive interaction is much stronger when two reactive sides are brought close to each other.\cite{sharifi2016pair} 
Furthermore, JPs were only observed to change their orientation at close distances and we did not observe significant long-range reorientations.
The JPs considered here are half coated, which even with a three-fold mobility difference, would yield a weak force-dipole hydrodynamic field\cite{popescu2018effective}: dominant flow field being $1/r^3$ i.e., source-dipole. Whereas, the chemical field decays more slowly as $1/r^2$. Furthermore, the momentum dampening of hydrodynamic effects might be exacerbated near boundaries exhibiting no-slip friction, while boundary-enhanced solute accumulation intensifies the repulsive chemical interactions.
Hence, the observations of direct contact between JPs, as illustrated in figure \ref{fig:collisions}, wherein JPs collide with their \ce{SiO2} hemispheres and subsequently move apart when their Pt sides come closer, support the expected chemophoretic interactions. 
Sharifi-Mood \etal predicted assembly of JPs in specific configurations,\cite{sharifi2016pair} which we do not observe in our experiments and is a key differentiation. 
One of the primary reasons behind this difference might be that their study doesn't account for mobility differences of the catalytic and inert halves that arise due to these halves having different interactions with solute molecules. Previous experiments\cite{campbell2019experimental} have shown that catalytic halves exhibit higher slip velocity and thus, higher mobility coefficient. 
Other possible contributing reasons behind this difference could be assumptions of neglecting Brownian fluctuations and variations of speeds in the colliding JPs.  We suspect that these differences allow the JPs to slide along each other and eventually detach in our experiments.

A recent semi-analytical study by \citet{nasouri2020exact} demonstrated that axisymmetrically interacting JPs can exhibit up to three fixed points, that determine the stability of distance between JPs. Their results for uniform mobility JPs were in close agreement with Sharifi-Mood \etal \cite{sharifi2016pair}, however, they found no fundamental addition of fixed points for the case of non-uniform mobility. They attributed this to mobility anisotropy only causing hydrodynamic modifications that are less important than chemical interactions. Although our current study does not indicate any stable fixed points, as all JPs eventually depart, we do observe a distribution of contact times that correlates with the various configurations of collisions. Furthermore, since most of the configurations are non-axisymmetric, we suspect that a significant role is played by asymmetry in mobility, which can introduce a chemical preference for the reorientation of JPs, during contact. Hence, in the next subsections, we characterize these configurations and analyze their kinematics of approach and departure.

\subsection{Classification of collisions}
\begin{figure*}[!ht]
\centering
\includegraphics[width=0.8\textwidth]{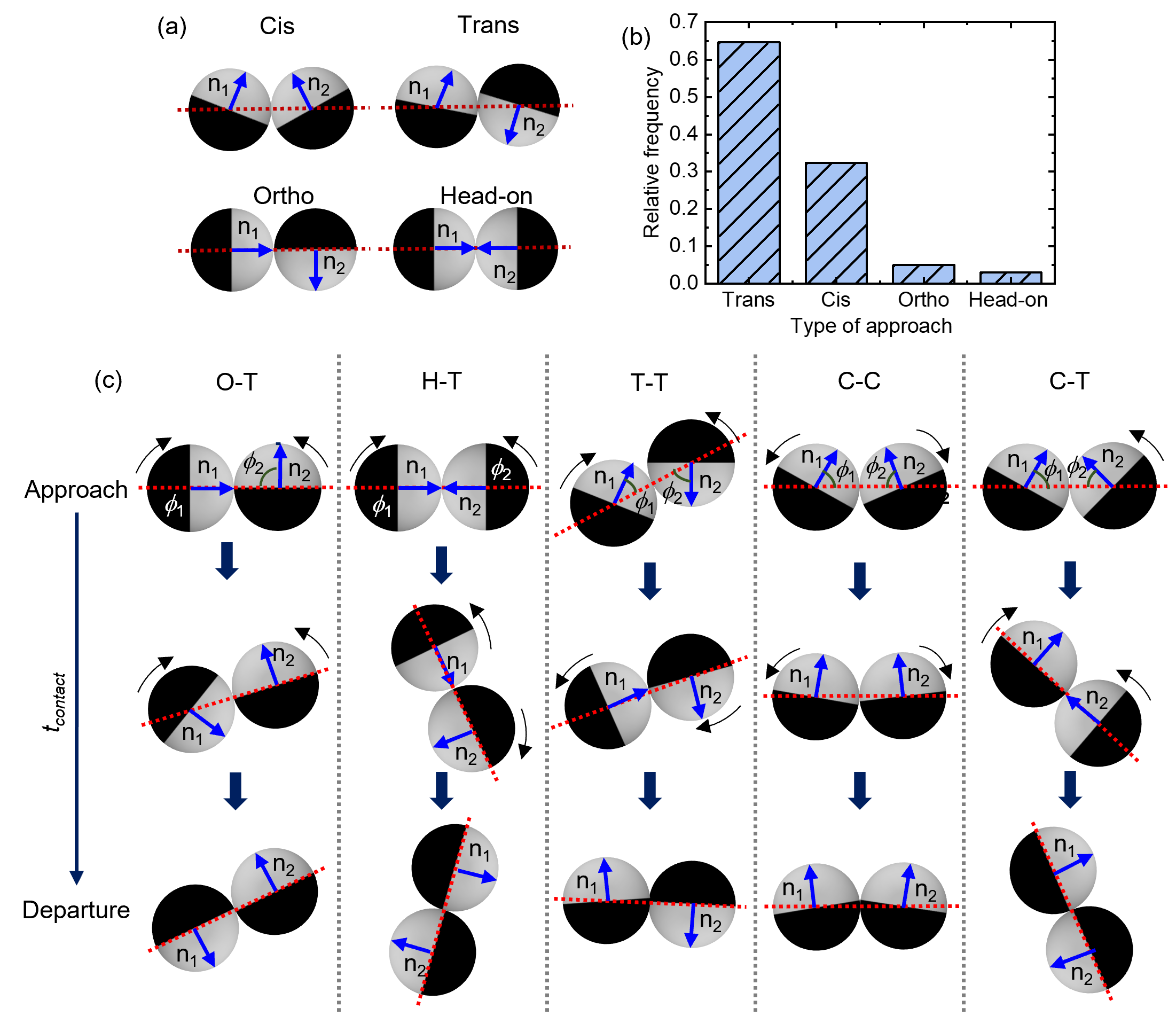}
\caption{(a) Schematic representing the different collisions observed. (b) Relative frequency distribution of the observed collisions. (c) Schematic depicting the sliding and rolling actions of the active JP pairs and the resulting orientation changes during various collision scenarios. The navy lines in the first row of figures indicate the rolling/sliding distance of the individual active JPs.}
\label{fig: collisiontype}
\end{figure*}

Each active JP undergoing a collision is characterized by its instantaneous speed $\textbf{v}_\text{inst.}$, and the orientation of a unit vector normal to the equator line of the JP, denoted by $\textbf{n}$. We broadly classify the collision into 4 categories by defining the orientation of JPs (i.e. $\textbf{n}{_{1}}$ and $\textbf{n}{_{2}}$) with respect to an imaginary line connecting their centers (as depicted in figure \ref{fig: collisiontype}(a)):
\begin{enumerate}[label=(\roman*)]
    \item \emph{Cis}: $\textbf{n}_1$ and $\textbf{n}_2$ are on the same side with respect to the centre-to-centre line.
    \item \emph{Trans}: $\textbf{n}_1$ and $\textbf{n}_2$ are on opposite sides with respect to the centre-to-centre line.
    \item \emph{Ortho}:  An intermediate between \emph{Cis} and \emph{Trans} states, where $\textbf{n}_1$ and $\textbf{n}_2$ are orthogonal to each other, with one of them oriented along the center-to-center line.
    \item \emph{Head-on}: Another possible intermediate state, where both $\textbf{n}_1$ and $\textbf{n}_2$ lie along the centre-to-centre line. It should be emphasized that here $\textbf{n}_1$ and $\textbf{n}_2$ always point in opposite directions, as collisions only take place between the \ce{SiO2} hemispheres.
\end{enumerate}

Figure \ref{fig: collisiontype}(b) shows the relative frequency of different approach orientations right before the collisions.
Our findings reveal that the likelihood of \emph{Trans} approach is almost twice that of \emph{Cis}.
Considering the repulsive nature of the chemical interactions due to solutes being generated at the Pt side, a higher likelihood of \emph{Trans} approach is expected because of the following reasons. 1. \textit{Ortho} and \textit{Head-on} are highly specific cases when normal vectors are very close being perpendicular and parallel, respectively, thus orientations generally fluctuate and form either \textit{Cis} or \textit{Trans} configuration. 2. As JPs approach each other, \textit{Cis} configuration experiences a higher chemical torque, as catalytic sides exhibit higher mobilities and also provide a chemically repulsive solute, yielding the largest chemical torque of all configurations, and thus, a lower probability of contact upon approach (further details and explanation provided in the next subsection).

Post-collision, most JP pairs depart in \emph{Trans} configuration, with some exception of \emph{Cis} departures, where JPs approach \emph{Cis} and detach from each other as \textit{Cis} with almost parallel $\textbf{n}_1$ and  $\textbf{n}_2$. These pure \textit{Cis} collisions occur rarely and constitute only $\sim$ 12.5 \% of the total \emph{Cis} collisions observed in this study. Therefore, we further refined the classification of collisions by additionally taking into account the orientations of the active JP pair while detaching from each other as (i) \emph{Ortho-Trans (O-T)}, (ii) \emph{Head-on-Trans (H-T)}, (iii) \emph{Trans-Trans (T-T)}, (iv) \emph{Cis-Cis (C-C)},  and (v) \emph{Cis-Trans (C-T)}. Here, the first and second designation indicates the approaching and detaching orientation of the JPs. Within the scope of this study, we did not observe other combinations such as \emph{T-C, O-C,} and \emph{H-C}. 

In figure \ref{fig: collisiontype}(c), we present a schematic representation of the interaction process for the active JPs pair during these five types of collisions. Overall, the orientation-dependent surrounding fluid flow and self-generated chemical fields dictate the sliding/rolling actions of the JP pair, which we will elaborate on in later sections. Meanwhile, the preliminary observations drawn from figure \ref{fig: collisiontype}(c) are that for an \emph{O-T} collision, the particle whose orientation vector is aligned parallel to the center-to-center line undergoes significant rotation as it encounters the other JP, as a result, the pair reorients to a \emph{Trans} configuration. In \emph{H-T} and \emph{T-T} collisions, the particles approach and roll in opposite directions till their equators are aligned, after which they detach in a \emph{Trans} configuration. The rare \emph{C-C} collisions are characterized by shorter rolling/sliding distances of the JPs. The approaching orientation of the JPs is such that their chemical fields are aligned, causing them to roll in opposite directions and depart in a \emph{Cis} configuration. In contrast, JPs undergoing \emph{C-T} collisions rotate in opposite directions, with one JP rolling more than the other and subsequently departing in a \emph{Trans} configuration.

\begin{figure*}[!t]
\centering
\includegraphics[width=\textwidth]{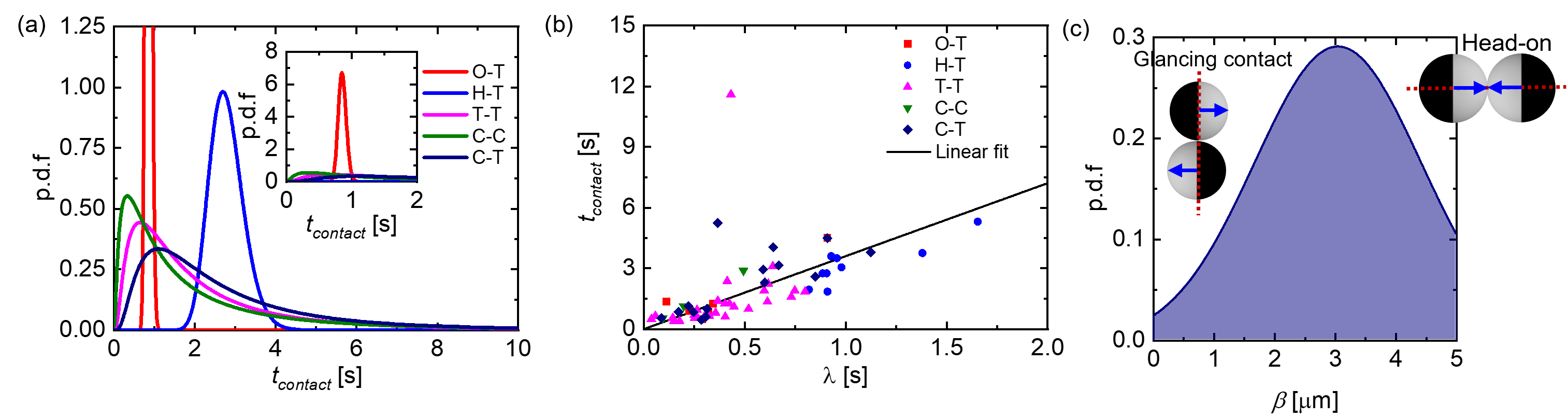}
\caption{(a) Fitted lognormal probability distribution curves of the contact time $t_\text{contact}$ for various collision scenarios classified based on the approach and departure orientations of the active JP pairs. (b) Variation of $t_\text{contact}$ with the approach parameter $\lambda$, and the corresponding linear fit (C-T collisions are excluded from the fit) (c) Fitted normal probability distribution curve of $\beta$. Inset depicts the two configurations at the extreme values of $\beta$.} 
\label{fig: Contact}
\end{figure*}

Figure \ref{fig: Contact}(a) displays the probability distribution of contact time ($t_\text{contact}$) for the observed collisions (see figure S2(a) in the Supporting Information for the raw data). \emph{T-T, C-C,} and \emph{C-T} exhibit a wide $t_\text{contact}$ distribution due to their multiple possible approach orientations, whereas \emph{O-T} and \emph{H-T} collisions have a narrow $t_\text{contact}$ distribution due to the very specific nature of their approach orientation. Among the broader distributions (\textit{T-T, C-C, and C-T}), \textit{C-T} spend a higher contact time on average, whereas \textit{C-C} collisions are relatively quicker. 

To obtain a simple estimate of $t_\text{contact}$ that takes into account both the approaching orientation and the JPs' speeds, we computed $\lambda = \beta/|{\textbf{v}_\text{relative}}|$, where $\beta = a(|cos\phi_1| + |cos\phi_2|)$, is the estimate for the contact length $l_\text{contact}$, $a$ is the radius of the JP and $\phi_1$ and $\phi_2$ are the angles between the center-to-center line and propulsion axis of JP$_1$ and JP$_2$, respectively, as depicted in figure \ref{fig: collisiontype}(c). Also, ${\textbf{v}_\text{relative}}=\textbf{v}_\text{inst.,1}-\textbf{v}_\text{inst.,2}$ just before the onset of the physical contact. Figure \ref{fig: Contact}(b) shows the comparison of experimentally measured $t_\text{contact}$ with estimated $\lambda$. Excluding \emph{C-T} collisions with higher $\lambda$, we find that $t_\text{contact} \propto \lambda$ with a proportionality constant $c=3.61 \pm 0.32$ indicating that $\lambda$ is underestimates $t_\text{contact}$. The linear relation between $t_\text{contact}$ and $\lambda$ indicates that in most of the collisions follow a simple unidirectional sliding/rolling of the JPs. Factors such as surface roughness, particle size polydispersity, and the thickness and uniformity of the Pt coating on the JPs may result in underestimating the time of contact, resulting in $c > 1$. For example, For \emph{H-T} collisions, it was also observed that the active JPs are locked for a few seconds at the beginning of collisions in the \emph{Head-on} state due to their opposing hydrodynamic interactions until a perturbation is induced by their thermal fluctuations, causing them to rotate and behave like JPs in a \emph{T-T} collision, such effects are not considered here. The positive deviation of the $t_\text{contact}$ values for \emph{C-T} collisions with high $\lambda$ (high $\beta$ and/or low $|\textbf{v}_\text{relative}|$) from the expected linear trend is expected to arise because of the rather complex sequence of steps during the physical contact of the JPs including intermittent switches of their rotation direction, which we will explain in a later section. Figure \ref{fig: Contact}(c) suggests that collisions with higher $\beta$ are more likely (see figure S2(b) in the Supporting Information for the raw data), which we shall also explain in later sections.

\subsection{Analyzing the kinematics of collision}

\begin{figure}[!ht]
\centering
\includegraphics[scale=0.8]{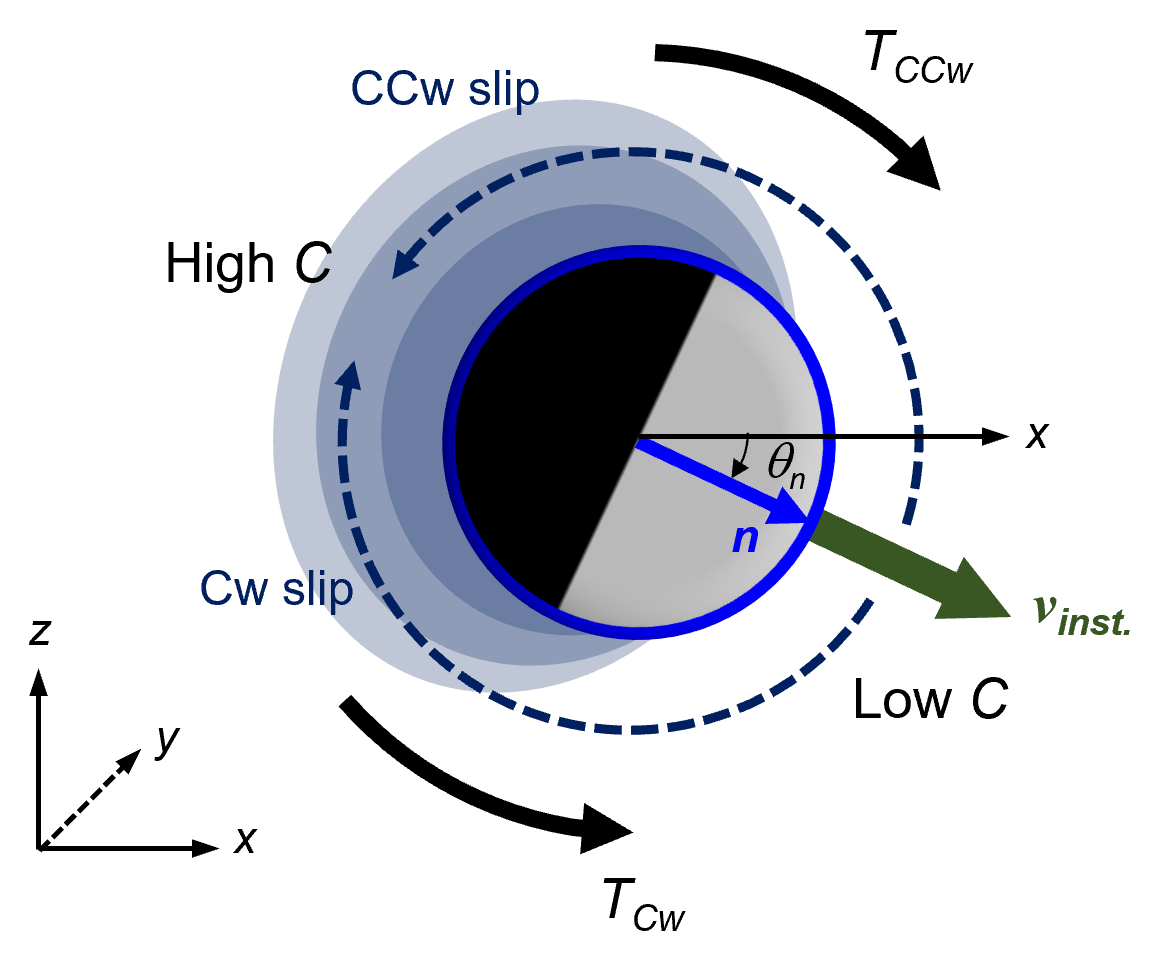}
\caption{Illustration of an isolated active JP experiencing phoretic slips and corresponding torques with no net rotation. Here $\theta_n$ represents the orientation of JP with respect to the lab frame X-axis.}
\label{fig: Balance}
\end{figure}

Next, we aim to understand how a specific type of pair collision impacts the translation and rotation of the JPs, as they approach and depart. As stated in the introduction section, although we observe a lack of far-field hydrodynamic interactions, the dynamics of collisions and contact are not simply steric but are determined by the chemical field distribution and phoretic slip-based interactions, which are elaborated below. As shown in the schematic shown in figure \ref{fig: Balance}(a), for an isolated \ce{SiO2}-Pt JP, the resultant surface (osmotic) slip due to the JP's self-generated chemical field causes it to propel forward. The clockwise (Cw) slip is known to generate a counterclockwise (CCw) torque T$_{CCw}$ about the propulsion direction \emph{i.e.} positive $z$-axis, and vice versa. Since these torques are equal and opposite, as expected, an isolated active JP does not experience net rotation due to propulsion \cite{mozaffari2016self}. 

When the JP approaches close to a solid boundary, such that the chemical field on the Pt side is not altered (in our case when a JP interacts with \ce{SiO2} side of other JP), the asymmetric viscous stress 
distribution of the surface slip can cause its rotation through a torque termed as the `propulsive torque' by Mozaffari \etal \cite{mozaffari2016self}. This propulsive torque tries to orient JP's \textbf{n} away from the boundary, which is different from the opposing viscous torque that tries to bring \textbf{n} towards the wall to roll the JP along the direction of motion parallel to the wall. On the other hand, if one JP interacts with an external asymmetric chemical field (in our case induced by another active JP), the mobility differences can yield a `chemical torque' \cite{golestanian2007designing,tuatulea2018artificial,vinze2021motion}. In the case of \ce{SiO2}-Pt, both catalytic and inert sides exhibit repulsive interactions with the released solute, yielding a slip from lower concentration to higher concentration region on the surface. However, the Pt side has been reported with significantly higher slip velocity\cite{campbell2019experimental}, suggesting an equally higher mobility on the catalytic side and rendering a reorientation such that the catalytic side faces furthest from the solute. During the pair-wise interactions, depending on the approach orientation of the JPs, both chemical and propulsive torques can contribute to the rotation of the interacting JPs, which we will discuss below through specific collisions. In addition to rotation, the collisions can also influence their translation motion through chemophoretic and hydrodynamic interactions, resulting in a change in the velocity $\textbf{v}_\text{inst.}$ of the JPs.

\subsubsection{The \textit{Ortho-Trans} collision} \label{sec:OT}

\begin{figure}[!ht]
\centering
\includegraphics[width=0.45\textwidth]{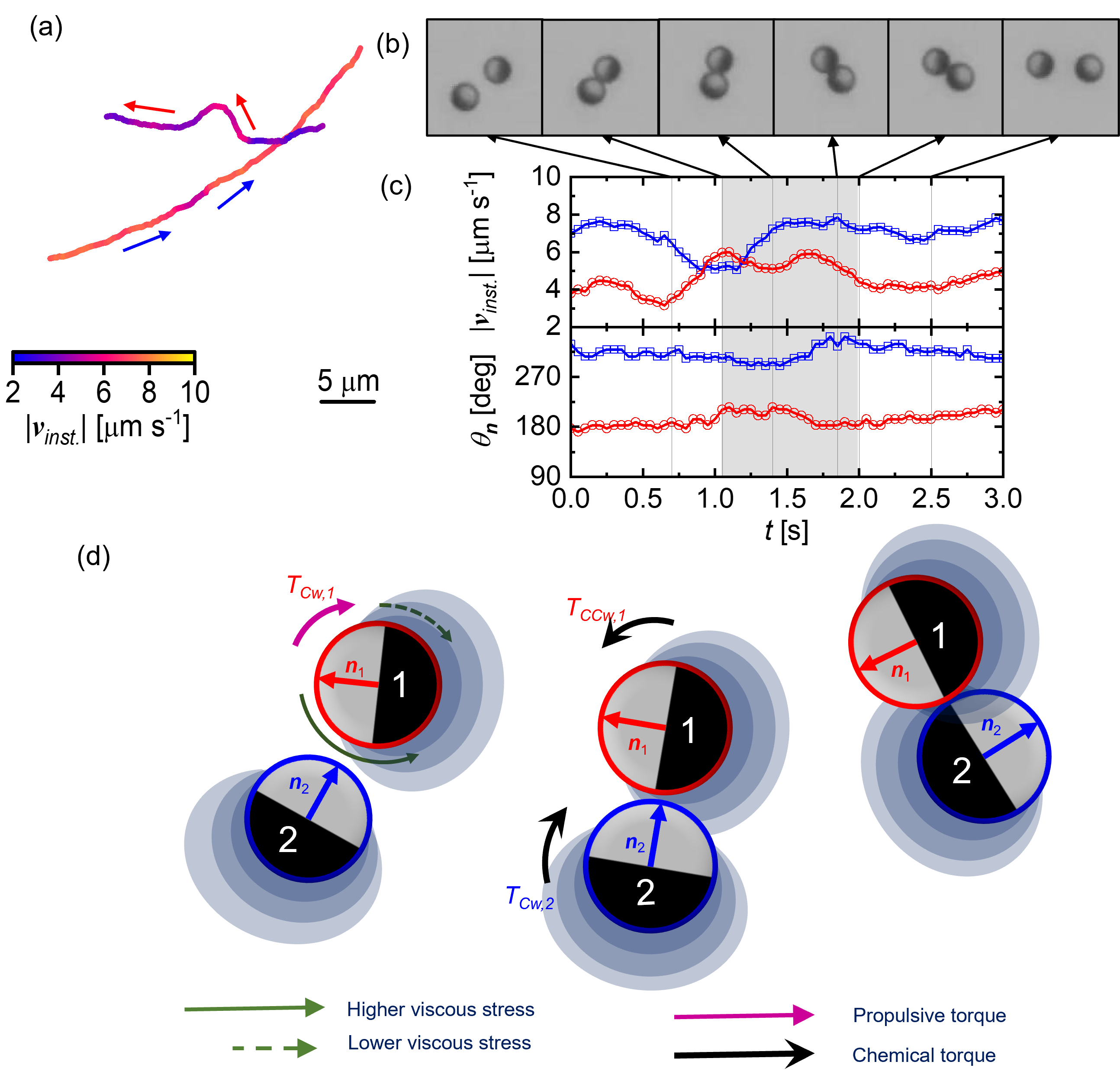}
\caption{For a representative \emph{O-T} collision ($\beta=2.93$ $\mu$m): (a) X-Y trajectories of the JPs, color-coded with instantaneous speeds. (b) Time-lapse micrographs demonstrating the sequence of JPs position and orientation (c) Variation of the instantaneous speed $|\textbf{v}_\text{inst.}|$ and instantaneous orientation ${\theta}_\text{\textbf{n}}$ for participating JPs participating with time. (d) Schematics demonstrating the rotational dynamics of JPs (left to right in time).}
\label{fig: O-T}
\end{figure}

In Figure \ref{fig: O-T}(a), we present the 2D trajectories of JPs participating in a representative \emph{O-T} collision ($\beta=2.93$ $\mu$m) (see supporting movie S2). Time-lapse images shown in figure \ref{fig: O-T}(b) demonstrate the JPs' position and orientation at different instances during the collision. The evolution of the instantaneous speed $|\textbf{v}_\text{inst.}|$ (computed with $\Delta t =$ 0.5 s to eliminate noise) and the orientation angle $\theta_\textbf{n}$ for the JP pair is demonstrated in figure \ref{fig: O-T}(c). On approach, just before the physical contact, the speed of the faster JP (2) decreases, while the JP (1) speeds up, and before detachment they return to their pre-collision speeds respectively.
Just before the impact, JP (1) first rotates CCw slightly, while JP (2) rotates Cw. This rotation allows the JPs to come in contact with an Ortho orientation. Subsequently, the direction of rotation is reversed before the eventual departure. 

To discern the aforementioned process, we first note that JPs approach each other in a near \textit{Ortho} manner. During this approach, the effect of far-field interactions remains weak, as affirmed by $\theta_\text{\textbf{n}}$ in figure \ref{fig: O-T}(c). Near contact, JP (1) experiences a Cw propulsive torque due to its bottom surface experiencing a higher shear in the narrow gap of \ce{SiO2} sides. Whereas, JP (2) experiences a weaker torque due to its surface slip being zero at the point of \textit{Ortho} contact i.e. along \textbf{n$_{2}$}. This is shown in the trajectories figure \ref{fig: O-T}(a) and schematic in figure \ref{fig: O-T}(d, left), where JP (2) experiences a very slight reorientation.
These rotations due to propulsive torque bring the JP pair to figure \ref{fig: O-T}(d, middle), where their exact \textit{Ortho} orientation experiences an opposing chemical torque due to the Pt side of JP (1). This chemical torque acts Cw on JP (2) and arises because of the lower mobility coefficient of \ce{SiO2} side; similarly, JP (1) rotates CCw. The rotation continues until the two Pt sides touch each other and the chemical repulsion of the solutes is strong enough to break the contact. Appendix (Section 5) provides further details on the estimates of chemical torque and how an \textit{Ortho} contact always experiences a chemical torque towards \textit{Trans} orientation. On the other hand, modifications in the translation motion of the JPs during the collision are rather straightforward. As JP (2) is more aligned with the center-to-center line, it experiences more steric hindrance due to JP (1), explaining the strong decrease in its speed. Whereas, a lesser alignment of JP (1) with respect to the center-to-center line allows JP (2) to gain some momentum by the incoming JP (1) resulting in its speeding up. After reorientation, once their equators align, JPs depart and gradually return to pre-collision speeds.

\subsubsection{The \textit{Head-on-Trans} and \textit{Trans-Trans} collision}

\begin{figure*}[!ht]
\centering
\includegraphics[width=\textwidth]{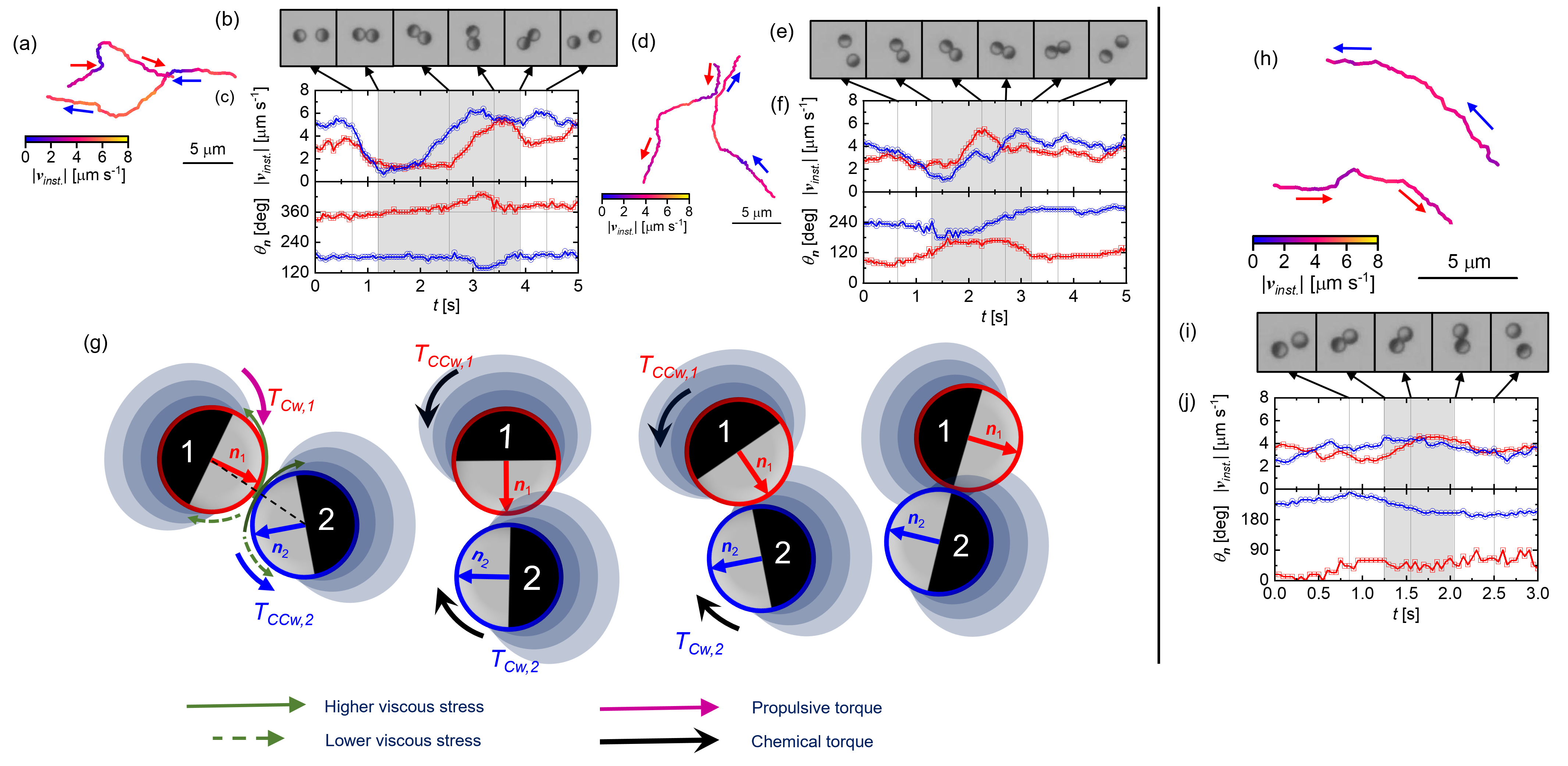}
\caption{For a representative \emph{H-T} collision ($\beta=4.93 \mu$m): (a) X-Y trajectories of the JPs, color-coded with instantaneous speeds. (b) Time-lapse micrographs demonstrating the sequence of JPs position and orientation (c) Variation of the instantaneous speed $|\textbf{v}_\text{inst.}|$ and instantaneous orientation ${\theta}_{\textbf{n}}$ for participating JPs participating with time. For a representative \emph{T-T} collision ($\beta=4.65$ $\mu$m) (d) X-Y trajectories of the JPs, color-coded with instantaneous speeds. (e) Time-lapse micrographs demonstrating the sequence of JPs position and orientation (f) Variation of the instantaneous speed $|\textbf{v}_\text{inst.}|$ and instantaneous orientation ${\theta}_{\textbf{n}}$ for participating JPs participating with time. (g) Schematics demonstrating the rotational dynamics of JPs of \emph{T-T} collision (left to right in time). (h) X-Y trajectories of JPs, color-coded with instantaneous speeds, participating in a representative \emph{T-T} collision with low $\beta(=1.81$ $\mu$m). (i) Time-lapse micrographs demonstrating the sequence of JPs position and orientation. (j) Variation of the instantaneous speed $|\textbf{v}_\text{inst.}|$ and orientation ${\theta}_{\textbf{n}}$.}
\label{fig: HTTT}
\end{figure*}

The behavior of $H-T$ and $T-T$ (high $\beta$) collisions (see representative movies S3 and S4 in Supporting Information) have a distinct response which we discuss through representative collisions shown in figure~\ref{fig: HTTT}.  Trajectories and temporal changes in JPs' position and orientation over the course of collision are shown in figure \ref{fig: HTTT}(a,c). Evidently, the JPs approach each other from their \ce{SiO2} sides, which ensures that the chemical interactions remain weak initially. The opposing hydrodynamic fields and lubrication effects\cite{popescu2018effective} substantially slow down both JPs during their approach, which recover to their original speed as they detach, as depicted in the evolution of the instantaneous speed $|\textbf{v}_\text{inst.}|$ for the two collisions. Furthermore, considering the higher $\beta$ (4.93 $\mu$m) for \emph{H-T} collisions in comparison with the lower $\beta$ (4.65 $\mu$m) for \emph{T-T} collision, the speed reduction is more pronounced in the former and recovery is also slow. 

As JPs try to maintain their propulsive momentum and move past each other, essentially in all \textit{Head-on} collisions, JPs first transition to the \textit{Trans} orientation. In such an orientation, the rotational behavior of JPs appears to be an intriguing two-step process, wherein, JPs first rotate to achieve an \textit{Trans} orientation. This is followed by a rotation in opposite directions to again achieve a \textit{Trans} orientation before detachment. The two-step process of \textit{T-O} and then \textit{O-T} is illustrated in figure \ref{fig: HTTT}(g). The \textit{T-O} transition is due to JP (2) experiencing a propulsive torque in the CCw direction. Consequently, JP (1) experiences a cog-like Cw rotation dominated by the slip of JP (2), as the propulsive torque JP (1) is weak due to its slip being weak at contact.
The subsequent \textit{O-T} transition follows the mechanism as discussed in the previous section \ref{sec:OT}.

Figure \ref{fig: HTTT} (h-i) shows\emph{T-T} collisions with low $\beta= 1.81 \mu$m. JPs undergo relatively lesser speed fluctuations and, importantly, only a single-step rotation under the influence of the mutual chemical field. In the limiting case where JPs approach in a grazing manner ($\beta \sim 0$ $\mu$m), they do not experience any significant speed reduction. 

\subsubsection{The \textit{Cis-Trans} and \textit{Cis-Cis} collision}
\begin{figure*}[ht]
\centering
\includegraphics[width=\textwidth]{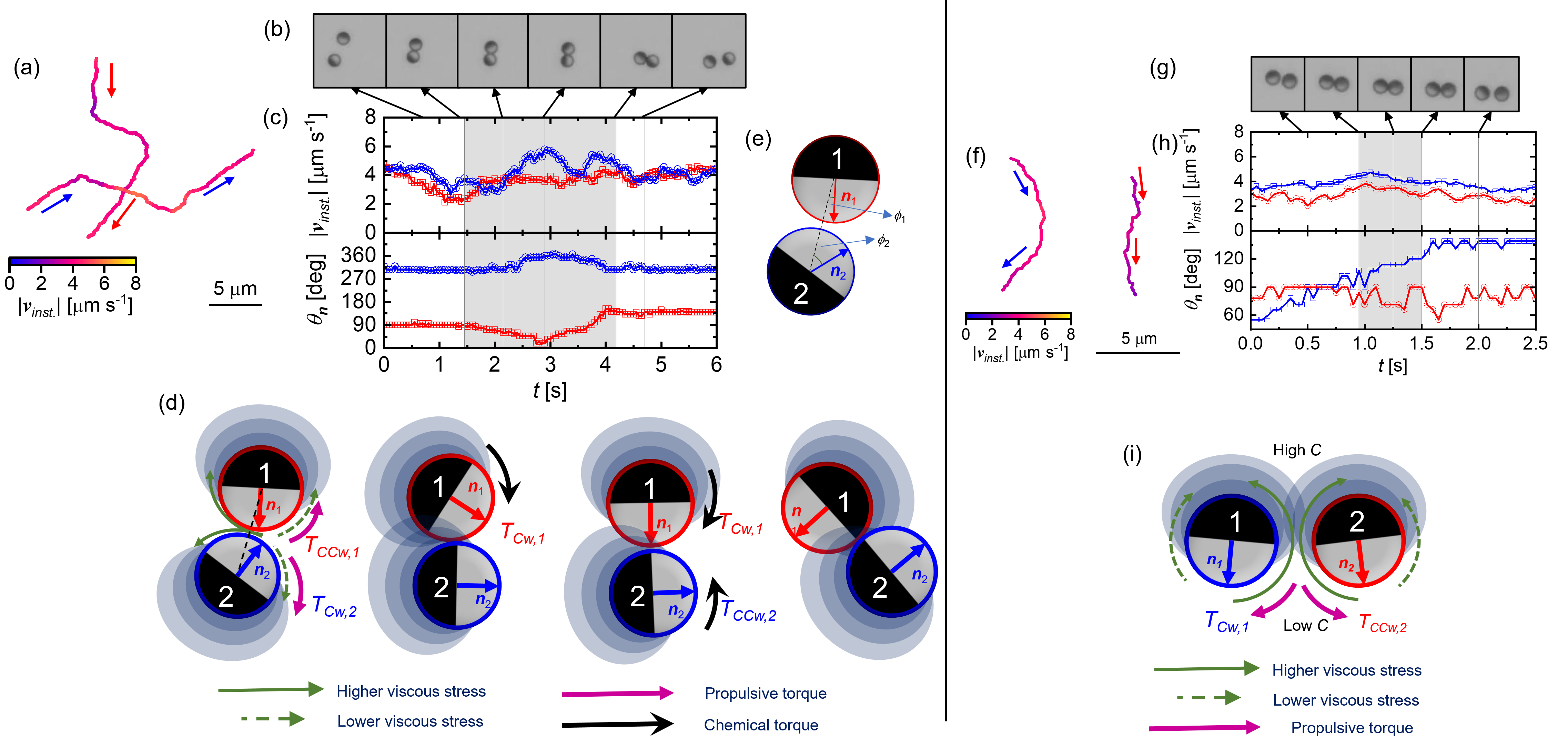}
\caption{For a representative \emph{C-T} collision ($\beta=4.28$ $\mu$m): (a) X-Y trajectories of the JPs, color-coded with instantaneous speeds. (b) Time-lapse micrographs demonstrating the sequence of the JPs' position and orientation. (c) Variation of the instantaneous speed $|\textbf{v}_\text{inst.}|$ and orientation ${\theta}_{\textbf{n}}$. (d) Illustration of the effect of phoretic slips on JPs in collisions, when two identical active JPs approach with a Cis orientation at high $\beta$, the differences in the surface flow on either side of the JPs result in a net propulsive torque, forcing them to rotate in specific directions. (e) Schematic illustration of the difference in alignment of the two JPs with respect to the center-to-center line, measured in terms of $\phi$. For a representative \emph{C-C} collision ($\beta=0.62$ $\mu$m): (f) X-Y trajectories of the JPs, color-coded with instantaneous speeds. (g) Time-lapse micrographs demonstrating the sequence of JPs position and orientation. (h) Variation of the instantaneous speed $|\textbf{v}_\text{inst.}|$ and orientation ${\theta}_{\textbf{n}}$. (i) Illustration of the effect of phoretic slips on JPs in collisions, when two identical active JPs approach with a Cis orientation at low $\beta$.}
\label{fig: CTCC}
\end{figure*}

Next, we discuss the case of \emph{C-T} collisions (see representative movie S5 in the supporting information). The trajectories and time-images shown in figure \ref{fig: CTCC}(a,b) demonstrate the JPs position and orientation at different instances during a representative collision. Figure \ref{fig: CTCC}(c) illustrate the corresponding evolution of $|{\textbf{v}_\text{inst.}|}$ and ${\theta}_{\textbf{n}}$ for the participating JPs. Unlike a \emph{C-C} collision (see figure \ref{fig: CTCC}), in this case, the JPs undergo more complicated speed and orientation changes. In this collision, despite a \emph{Cis} approach the orientation vectors of the JPs are significantly misaligned such that there is always a finite approaching velocity along the center-to-center line at the time of first contact. As a result, at the contact, the two reactive Pt sides remain distant, and primary interactions are mostly steric, resulting in an initial slow-down of the JPs. Immediately after the contact, the initial rotation is dictated by the altered propulsive torque distribution due to the thin gap between the \ce{SiO2} sides of the JPs. This aligns the equators of the JPs bringing their reactive Pt sides closer. Such an orientation provides an enhanced driving push due to their combined chemical field initially enhancing the speeds of the participating JPs. It was also observed that the alignment of the two JPs with respect to the center-to-center line, measured in terms of $\phi$ (see schematic shown in figure \ref{fig: CTCC}(e)) was not the same ($\phi_{1}(= 15$\textdegree$) < \phi_{2}(= 25$\textdegree)). Due to this orientation difference, we believe that JP (1) experiences more resistance and thus undergoes more speed reduction during the approach and slower acceleration (both translation and rotation). This competition between the JPs brings JP (1) to the offside front of JP (2). The chemical field of JP (1)'s Pt side now starts to affect the rotation of JP (2), forcing a Cw chemical torque, bringing the JPs to an \emph{Ortho} orientation followed by \emph{Trans} separation (as discussed in section \ref{sec:OT}). The apparent direction reversal maneuver performed by JP (1) results in an unpredictably larger $t_\text{contact}$ (see figure \ref{fig: Contact}(b)). 

Using figure \ref{fig: CTCC}(g-i), we finally discuss the behavior of a JP pair undergoing a \emph{C-C} collision (see representative movie S6 in the supporting information). Time images shown in figure \ref{fig: CTCC}(g) demonstrate the JPs' position and orientation at different instances during the collision. While approaching, the JPs align with their normal vectors being nearly parallel. As evident from the evolution of the instantaneous speed $|{\textbf{v}_\text{inst.}|}$ of the participating JPs (figure \ref{fig: CTCC}(h)), such an orientation provides an enhanced driving push due to their combined chemical field initially enhancing the speeds of the participating JPs. Upon detachment, their speeds almost return to their pre-collision values. This observation is consistent with recent numerical simulations \cite{sharifi2016pair}. \\

The proximity of the Pt sides suggests that chemical torques should force rotate the JPs to attain a \textit{head-on} orientation. Also, this effect is expected to be stronger than the \textit{Ortho} case (refer to Appendix). However, the rotation of JPs appears to be consistent with the propulsive torques. Although this anomalous behavior is not very clear, one possible reason could be the increased strength of propulsive torque due to the increased slip in the gap between the JPs. Nonetheless, the occurrence of such collisions is very low.

\subsection{Non-contact near-field interactions}

\begin{figure*}[ht]
\centering
\includegraphics[width=0.75\textwidth]{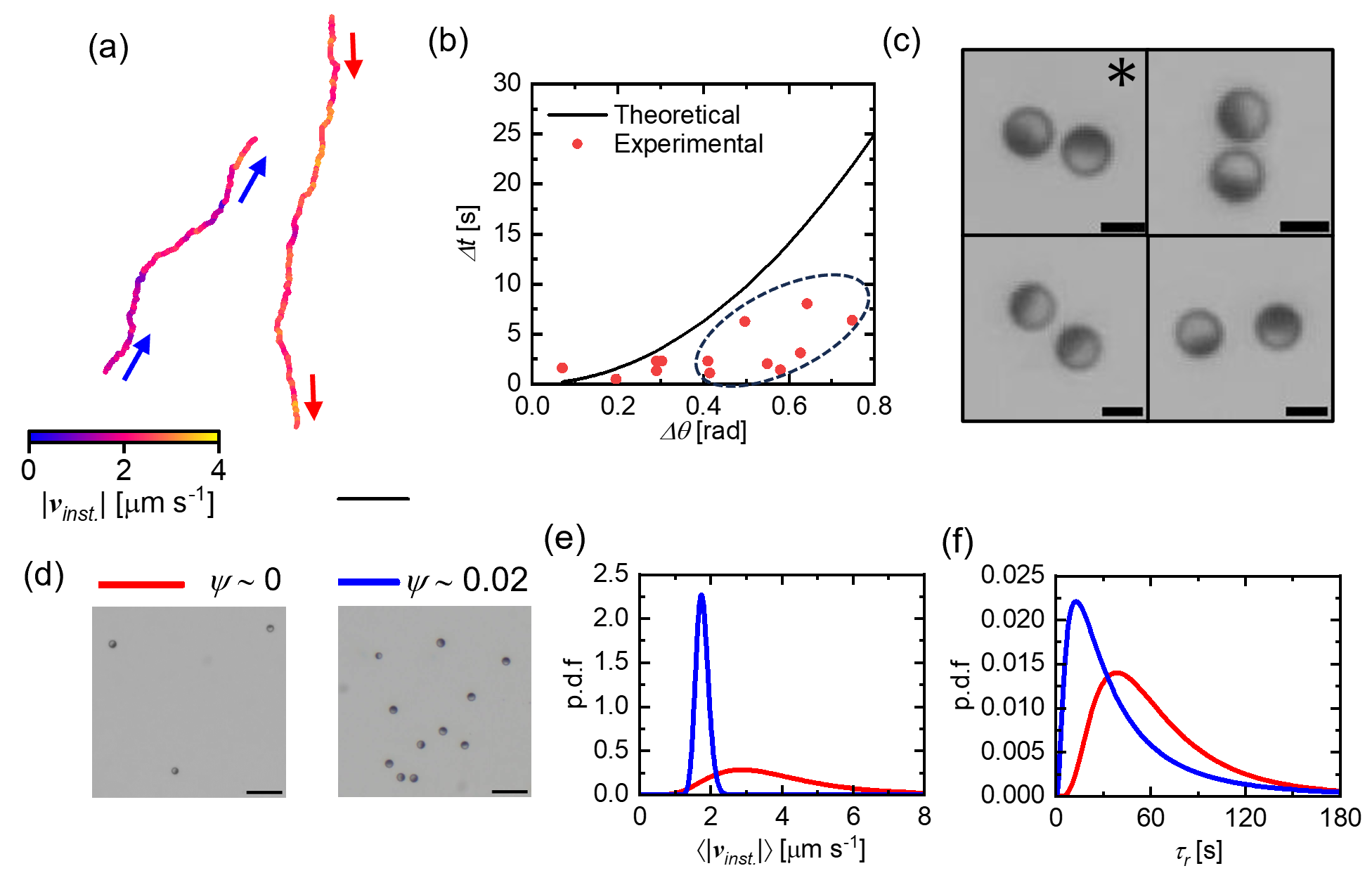}
\caption{(a) Representative trajectories ($\sim$ 15 s) of two active JPs participating in a \emph{non-contact} interaction. The scale bar indicates a length of 5 $\mu$m. (b) Comparison of theoretical (Stokes-Einstein) and experimental time scales required for deflection. (c) Representative optical micrograph images of separate \emph{non-contact} interactions at the instant of the least inter-particle distance $d_\text{cc, min.}$. The JP pair marked with an asterisk corresponds to the trajectory shown in Figure (a). The scale bars indicate a length of 5 $\mu$m. (d) Optical micrographs depicting a dilute (top) and a populated (bottom) system of active JPs. The scale bars in the optical micrographs indicate a 25 $\mu$m length. (e,f) Fitted log-normal probability distribution curves of the average instantaneous speed $\langle|\textbf{v}_\text{inst.}|\rangle$ and reorientation timescale $\tau_\text{r}$ for systems with different particle area fraction $\psi$.}
\label{fig: nc}
\end{figure*}

In addition to the physical collisions, we also observed events where active JPs interact despite being physically distant, i.e., $ 5 {\mu}m < d_\text{cc} < 15 {\mu}m $ (see Movie S7 in the Supporting Information). Figure \ref{fig: nc}(a) displays the trajectories of two active JPs engaging in one such \emph{non-contact} interaction. To distinguish these interactions induced orientation change with those caused by the Brownian fluctuations, in figure \ref{fig: nc}(b) we show the $\Delta t$ values representing the time taken to undergo the measured orientation change $\Delta \theta$ for several pairs of JPs maintaining $d_\text{cc}$ $>$ 5 $\mu$m all the time. The experimentally measured values are compared with the theoretically estimated values $\sim{\frac{\langle\Delta \theta^2\rangle}{2D_\text{r}}}$ (solid line). Here, $D_\text{r}$ is the experimentally measured rotational diffusivity of isolated JPs. Interactions, where the theoretically estimated values are greater than the experimentally measured values (encircled), represent the orientation fluctuations being caused by the \emph{non-contact} interactions. Furthermore, another key observation is that in all such interactions, the Pt side of at least one of the JPs is oriented towards the other JP to facilitate the chemical interactions. Figure \ref{fig: nc} (c) shows the optical micrograph images for a few representative \emph{non-contact} interactions when the JPs are at their closest, corresponding to the least interparticle distance $d_\text{cc,min.}$. While such interactions can be long-ranged and span for distances up to multiple particle diameters, we did not observe any noticeable change in ${\theta}_{\textbf{n}}$ for an active JP that could be potentially caused by the other JP beyond an $d_\text{cc,min.}$ $\sim$ $3d$ (i.e., 15 ${\mu}m$). This affirms the negligible role of far-field hydrodynamic interactions in governing the pair interactions (contact or non-contact) of self-propelled \ce{SiO2}-Pt JPs.\\

An increase in the population of active JPs in the system leads to a higher occurrence of multi-particle interactions ($n > 2$). Understanding pairwise interactions of isolated active JP pairs is a crucial aspect of comprehending the mechanics of the crowded active systems and their dynamic assemblies, which generally germinate with the collisions of two JPs. Having established that an individual active JP generally undergoes reorientation and speed reduction while physically interacting with a similar active JP, it is reasonable to assume that the consequences would be more pronounced for relatively denser systems. 
To validate this, we compare the distribution of $\langle|\textbf{v}_\text{inst.}|\rangle$ and $\tau_\text{r}$ values of active JPs in a relatively densely populated system (particle area fraction $\psi \sim 0.02$, see optical micrograph figure \ref{fig: nc}(d)) to that of a dilute system ($\psi \sim 5 \times 10^{-4}$, see optical micrograph figure \ref{fig: nc}(d)). As shown in figure \ref{fig: nc}(e) and figure \ref{fig: nc}(f), as expected, we find that the peak positions of both $\langle|\textbf{v}_\text{inst.}|\rangle$ and $\tau_\text{r}$ decrease with an increase in $\psi$ due to the increased inter-particle interactions. Also, with an increase in $\psi$, we find that the peaks of probability distribution curves narrow down, suggesting that the behavior of particles in the system is almost identical, indicating an onset of dynamical ordering in the system (see figure S2(c) in the Supporting Information for the raw data of $\langle|\textbf{v}_\text{inst.}|\rangle$ and $\tau_\text{r}$). 
The uniformity in the particles' behavior is likely the coarse-grained effect of the increase in pairwise and multi-JPs interactions (both physical and far-field) in a system with largely evenly distributed JPs in the 2-D space. Further insights into the collective response can be achieved through more dedicated experimental investigations.

\section{Summary}
In this study, we experimentally study the pair-wise interactions of \ce{H2O2} fuelled active \ce{SiO2}-Pt JPs of identical size. It was observed that in almost all collisions, the JPs approach with their \ce{SiO2} hemispheres coming in contact first. Subsequently, they rotate (about an axis perpendicular to the bottom wall) and slide along each other eventually aligning their Janus planes, mostly from from opposite sides. 
On achieving such orientation JPs detach from each other. 
Based on the approach orientation of JPs' normal vectors with respect to the center-to-center line, the collisions have been broadly classified as \emph{Cis, Trans, Ortho, and Head-on}.  
The interaction duration and the overall impact of the collision are contingent upon the approach orientation and the velocities of the active JPs involved. In some instances, even without physical contact, JPs are influenced by another approaching JP, and such interactions are designated as \textit{non-contact} collisions.

\begin{figure*}[ht]
\centering
\includegraphics[width=\textwidth]{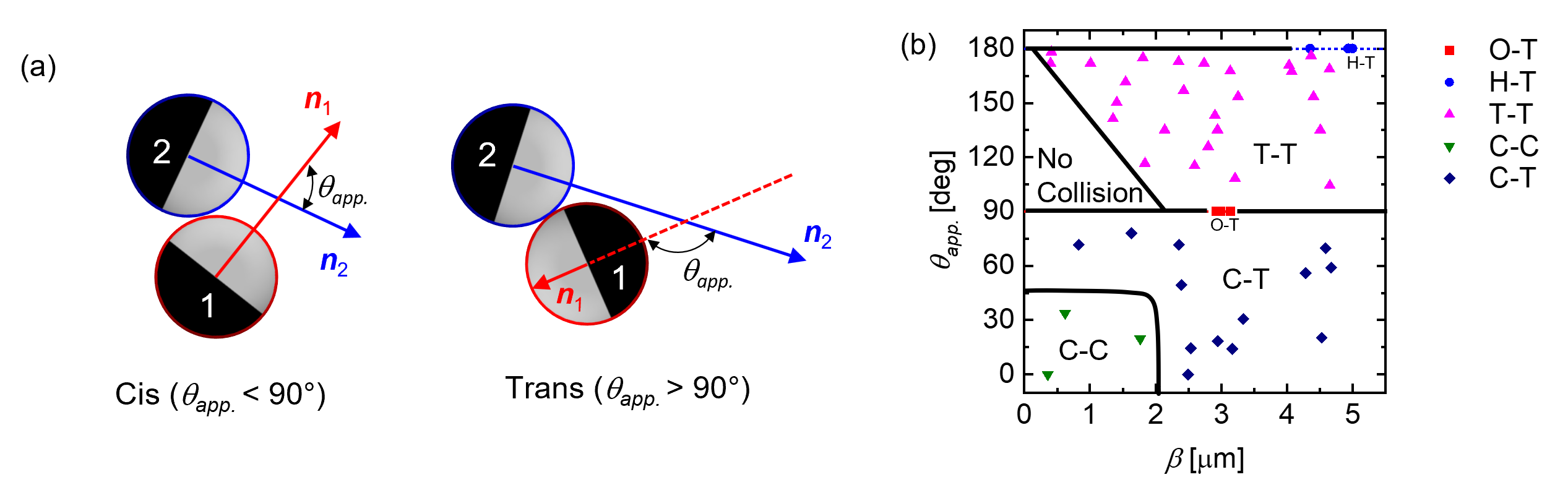}
\caption{(a) Schematics illustrating the approach angle $\theta_\text{app.}$ for representative \emph{Cis} and \emph{Trans} orientations. (b) Phase diagram based on the approach angle $\theta_\text{app.}$ and the overlap of JPs captured by $\beta$.}
\label{fig:Phased}
\end{figure*}

Based on our observations from all different kinds of pair-interactions a phase diagram has been constructed (see figure \ref{fig:Phased}). The plot serves as a useful guide in predicting the nature of the collision and thus the interaction outcome, based on the approach orientation quantified in terms of angle $\theta_\text{app.}$ (Y-axis) and the overlap of JPs captured by $\beta$ (X-axis). Here low $\beta$ represents glancing contact and high $\beta$ represents \textit{Head-on} like collisions. 

In accordance with the established behavior of a chemically active JP near a solid wall, the collision dynamics (both rotational and translation) was found to be affected by the nearby JPs. This influence stems from the alteration of the propulsive torque (due to viscous stress associated with phoretic slip) and/or the chemical torque acting on the JP. When the approach orientation positions the reactive Pt side of a JP in the vicinity of the \ce{SiO2} (\emph{i.e. Ortho}) or the Pt side (\emph{i.e. Cis}) of the other JP, the interactions are mostly influenced by the distribution of the chemical fields. 
In $Head-on$ and $Trans$ approaches, since the reactive Pt sides of the JPs remain oriented away from each other, 
during the approach the interactions are mostly influenced by the modification in the propulsive torques. This rotation brings the JPs in $Ortho$ orientation, initiating the influence of chemical fields, eventually separating with $Trans$ orientation. To this end, we did not observe any concrete evidence of far-field hydrodynamics to play a significant role in the pair interactions. The effect of approach orientation on the onset of chemical interactions is further supported using a simple qualitative estimate of the chemical torques on the JPs, which is detailed in the appendix section. 
Comprehending the pairwise interactions among active JPs stands as a crucial cornerstone in our efforts to grasp the intricate dynamics of JPs within densely populated environments characterized by multiple-body interactions, which we have briefly discussed towards the end of section 3.5. 

Additional experiments are needed to further understand the onset of dynamic assemblies because the pair-wise investigation presented here and the one reported by Sharan \textit{et al.} reports an eventual scatter, but numerous experiments outlined in the Introduction section report assemblies and dynamical structures \cite{sharan2023pair}. This might be due to the use of \ce{Au-Pt} particles that operate predominantly via self-electrophoresis: electrochemical decomposition of \ce{H2O2} generates electrostatic fields that act on top of the chemical and hydrodynamic fields, probably resulting in enhanced attraction between the JPs \cite{Anderson1989}. Indeed, pair-interaction studies on such rod-like particles have shown an array of robust pair-wise assemblies \cite{wykes2016dynamic,jewell2016catalytically}. 
To the best of our knowledge, such isolated pair-wise interactions are not yet examined for spherical \ce{Au-Pt} JPs. Understanding their assembly in contrast to rod shaped JPs can facilitate ideas for unique functionality because their sphericity can offer higher symmetry \& isotropy in dynamical structures than those made of rods (which may facilitate more anisotropic structures); furthermore, spherical shapes offer the highest surface area. 
Additionally, studying multi-body interactions of the current \ce{Pt-SiO2} particles may also introduce additional features that can lead to clustering, as was also shown theoretically by \citet{varma2019modeling}: three-body interactions imparts an additional drift to JPs that is absent in pairwise interactions. We shall address these aspects in our upcoming study.

While our study provides conclusive experimental evidence about the dominant role of chemical interactions in governing the pair-encounters of \ce{SiO2}-Pt JPs, the exact quantification of the competingand coupled chemo-hydrodynamic 
effects is require dedicated simulation studies. However, we believe that upon changing the fuel concentration, thickness of Pt coating, confinement of optical cell, and particle size disparity, the relative strengths of the two interactions can be tuned resulting in a modified response. In addition, by altering the mode of swimming by either changing the active surface coverage \cite{popescu2018effective, sharifi2016pair}, or altering the surface chemistry of the JPs, \cite{sharan2023pair} distinct yet connected response can be observed.


\newpage

\section{Appendix: Estimating the chemical torque}
\begin{figure*}[ht]
		\centering
		\includegraphics[width=0.75\textwidth]{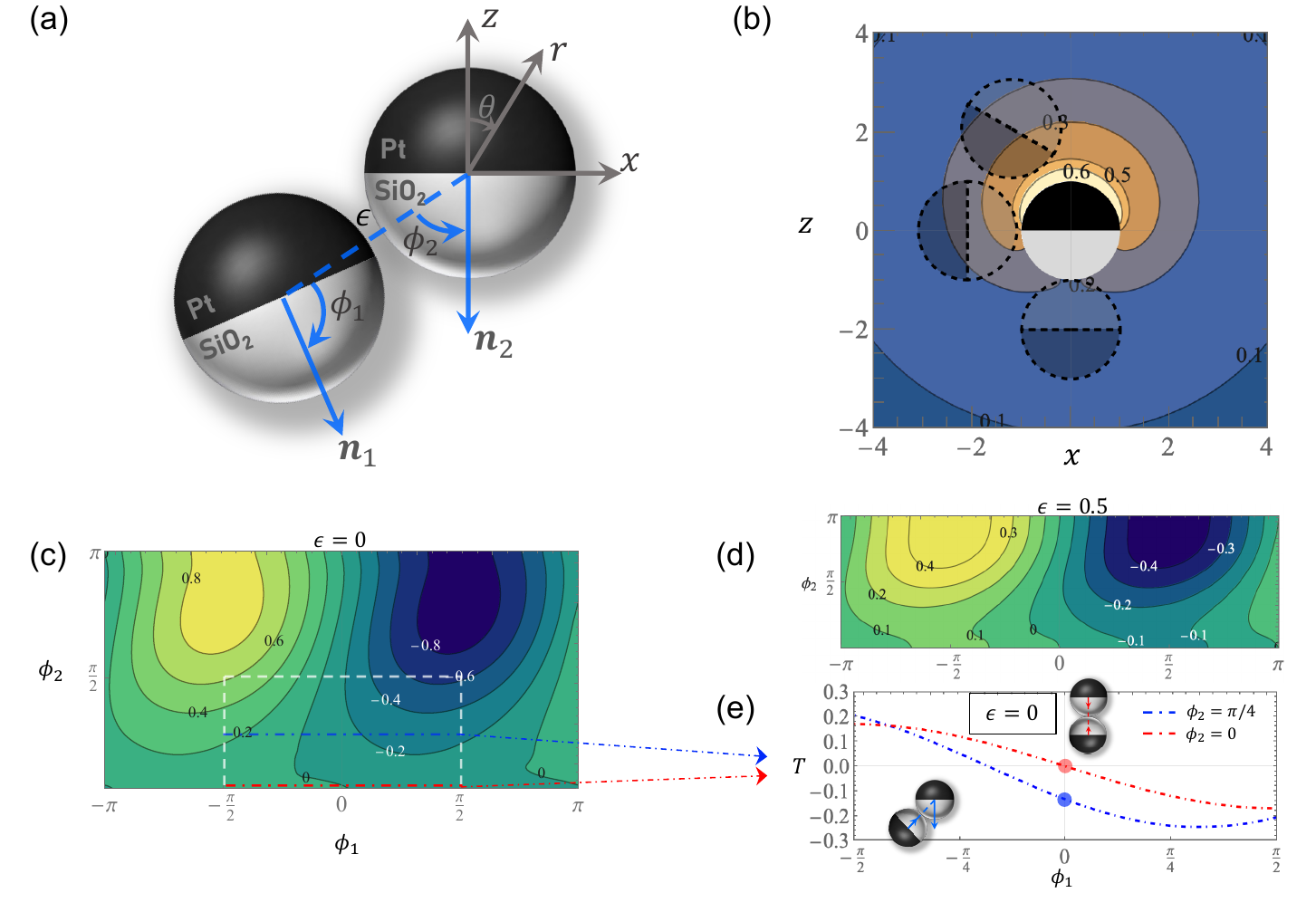}
		\caption{(a) Schematic of two Janus particles (JP) interacting chemically that yields an instantaneous chemical torque. The angles from center-to-center dotted line to the propulsive axis of JP\textsubscript{1} and JP\textsubscript{2} form $\phi_1$ and $\phi_2$ angles, respectively. The length of center to center line is $2 a^{*} +\epsilon^{*}$, where $\epsilon^{*}$ is the minimum gap between two JPs and $a^{*}$ denotes the particle radius (with * representing dimensionality).
			(b) Illustration of various configurations of JP\textsubscript{1} around a JP\textsubscript{2} that is fixed with $\IB{n}_2 = -\IB{e}_z $. To explore all possible cases, we consider $\phi_2 \in [0,\pi] $ and $\phi_1 \in [-\pi,\pi]$; three example cases are shown of \textit{Trans, Ortho}, and \textit{Head-on}. Here we opt for a right-handed coordinate system with the $ y- $axis pointing into the plane.
   (c) Chemical torque (T) on JP$_1$ due to JP$_2$ at $\epsilon^{*}=0$ and (d) $\epsilon^{*}= 0.5 a^{*}$ (dashed white lines depict the region of collision configurations realized in the current experiments). 
   (e) Chemical torque on JP$_1$ for various tilts ($\phi_1$) at two contact points with JP$_2$ ($\phi_2=\pi/4$ and 0), shown by dashed dot lines in (c).
   The schematics embedded in (e) represent the configuration of blue and red points for $\phi_2=\pi/4$ and $\phi_2=0$, respectively.}%
		\label{fig:contour}%
	\end{figure*}

Results in the main text show that in certain orientations where the Pt sides are oriented away, the chemical interactions remain weaker. 
It is when the orientations are such that the Pt side of either active JP is in the vicinity of another JP, the chemical interaction-induced reorientations become substantial. To support this argument, for a pair of active JPs (see schematic shown in figure \ref{fig:contour}) we performed simple calculations on obtaining qualitative insights on the chemical torque induced from mobility anisotropy, which implicitly assumes that a. self-interactions are weaker than those imposed by JP$_2$ on JP$_1$ (or vice-versa), and b. chemical field from JP$_2$ acts as linear gradients across JP$_1$. 

Considering only the chemical impact of one JP on the other, we first calculate the solute field around $ \text{JP}_{2} $, which points to its propulsion vector ($ \IB{n}_{2} $) in the negative $ z $-axis.
 The self-propulsion speed ($ U^{*} $) of a Janus particle of size ($ a^{*}  $) $ \sim 5 \mu $m is $  \sim 2-6 \times 10^{-6} $ m/s. For solute diffusion coefficient  $ D^{*}_{\text{solute}} \sim10^{-9} $ m$ ^{2} $/s, the associated  Péclet number ($\mbox{Pe} = U^{*} a^{*}/D^{*} $) is small $ \sim 10^{-2} $, and thus we neglect the advective effects.
Furthermore, we assume that the system is quasi-steady in the concentration field: the time scale of reorientation is larger than that of solute diffusion around the particle ($ a^{* \, 2} /D^{*}_{\text{solute}} \sim  10^{-2} s $).
Consequently, the disturbance concentration field around $ \text{JP}_{2} $ is governed by the Laplace equation; the boundary conditions are governed by a step flux at the particle surface and a decaying condition at the infinity:

\begin{subequations}
    \begin{gather}
		\nabla^{2} c = 0, \label{GE:conc}\\
		\left.\PD{c}{r}\right\vert_{r=1}
		= \mathcal{A}(\theta) = \left\{
		\begin{array}{ll}
			\mathcal{A}_{+} = -1 \qquad \theta \leq \theta_{c}\\
			\mathcal{A}_{-} = 	0 \qquad \theta > \theta_{c}
		\end{array}	\right. 
		\; \mbox{and} \\
  \quad c\rightarrow0 \;  \; \mbox{as } r \rightarrow \infty.
		\label{BC:conc}
	\end{gather}
\end{subequations}

The negative sign in the boundary condition indicates that the solute gradient decreases as we move away from the reactive side.
In the above equations, length and concentration are non-dimensionalized using $ a^{*} $, $ |\mathcal{A}^{*}|a^{*}/{D^{*}} $. Here, $  |\mathcal{A}^{*}| $ is the maximum magnitude of dimensional activity in the units of $ \text{M} \text{m}^{-2} \text{s}^{-1} $.
Following Golestanian \etal \cite{golestanian2007designing}, we obtain concentration field around $ \text{JP}_{2} $  as
\begin{equation}\label{Conc}
	c(r,\theta)=\displaystyle\sum_{n=0}^{\infty} {\frac{-\mathcal{A}_{n}}{(n+1)}}\, \frac{P_{n}(\cos \theta)}{r^{n+1}},
\end{equation}
where $ P_{n} $ is the $ n $th order Legendre polynomial and $ \mathcal{A}_{n} $ are the coefficients of activity distribution: $ 	\mathcal{A}(\theta)=\sum_{n=0}^{\infty} \mathcal{A}_{n} P_{n}(cos\theta) $.
These coefficients are found by taking an inner product with the Legendre polynomials and are obtained as
\begin{align}\label{A}
	\mathcal{A}_{0} &=\frac{(1-\cos \theta_{c})}{2} \; \mbox{ and\ } \nonumber \\  \;  \mathcal{A}_{n} &=\frac{-1}{2} (P_{n+1}(\cos \theta_{c}) - P_{n-1}(\cos \theta_{c})) \; \mbox{for\ } n \geq 1.
\end{align}
In such a self-generated solute field, a JP will translate due to slip on its surface: $ \mathcal{M} \bnabla_{s} c $, where $ \bnabla_{s} $ is the surface gradient vector and $ \mathcal{M} $ is the surface mobility of solute, non-dimensionalized by $ \mathcal{M}^{*} $ ($ \frac{k_{B} T a_{\text{sol}}^{* \, 2}}{2 \mu} \sim 10^{-32} \, \text{m}^{5} \text{s}^{-1} $) \cite{Anderson1989}. 
The mobility coefficient is generally positive for both Pt and SiO$ _{2} $ (depicting repulsive solute-surface interactions). This slip is axisymmetric (only having $ \IB{e}_{\theta} $ component) and can not generate a rotation velocity even if mobility differences between the two halves exist. 
In addition to the anisotropic mobility distribution, the particle needs to experience a solute field that is not axisymmetric.
Pair interactions with another JP break this symmetry, yielding a chemical torque, which imparts a rotation velocity to the JP in order for it to remain torque-free.

We now evaluate the torque experienced by JP$ _{1} $ from the solute field of $ \text{JP}_{2} $ (\ref{Conc}) at a non-dimensional distance of $ 2+\epsilon $, where $ \epsilon $ is the minimum gap between the particles (such that $ \epsilon \leq 0.5 $). 
The primary torque experienced would be due to the mobility difference at the two halves ($ \mathcal{M}_{+} \neq \mathcal{M}_{-} $). For Pt-SiO$_{2} $, Campbell \etal showed a slip anisotropy on the two halves of the particle, suggesting a mobility difference of up to a factor of 3 \cite{campbell2019experimental}.
Here we consider the mobility anisotropy as $ \mathcal{M}_{+} = 3 \mathcal{M}_{-} = 3 $.
First, we evaluate the chemical gradient around an imaginary JP$ _{1} $ as shown in figure \ref{fig:contour}(b). 
We shift the coordinates from the centre of JP$ _{2} $ to JP$ _{1} $ by using the following transformation as replacement: $ z_{2} = z_{1} - (2+\epsilon) \cos \phi_2, \;  x_{2} = x_{1} - (2+\epsilon) \sin \phi_2 $. 
These coordinates are finally rotated to account for the tilt of JP$ _{1} $ with respect to negative $ z $-axis ($ \Phi = \pi - \phi_1 - \phi_2 $): $ z_{1} = x \cos \Phi - z\sin \Phi, \;  x_{1} = x \sin \Phi + z\cos \Phi $.
We then estimate the gradient across this imaginary JP$ _{1} $ by evaluating the solute concentration at the $ r=1 $. This provides us with maximum and minimum concentration and their respective $\theta$, which helps us formulate the passive diffusiophoretic problem in the external gradient $\left( \gamma = (c_{\text{max}}-c_{\text{min}})/2 \right)$ of JP$ _{2} $. 
This passive diffusiophoresis problem yields the solution as\cite{tuatulea2018artificial,vinze2021motion}:
\begin{equation}
\tilde{c} =	\gamma \frac{z \cos \Theta + x \cos \Theta}{(x^{2}+y^{2}+z^{2})^{3/2}},
\end{equation}
where $ \Theta $ is the angle directing the gradient from minimum to maximum value across JP$ _{1} $;
$\tilde{c}$ is the approximate solute concentration field around JP$ _{1} $ due to an external gradient from JP$ _{2} $. This field is not axisymmetric and can induce a chemical torque if $ \mathcal{M}_{+} \neq \mathcal{M}_{-} $ \cite{golestanian2007designing}.
The instantaneous chemical torque at each configuration of JP$ _{1} $ is evaluated using the expression derived by \citet{stone1996propulsion} for microswimmers:
\begin{align}
	\IB{T}  = & - \int_{S} \IB{e}_{r} \times \IB{u}_{\text{slip}} \, \text{d}S = - \int_{S} \IB{e}_{r} \times  [ \mathcal{M}(\theta) \bnabla_{s} \tilde{c} ]  \, \text{d}S \nonumber \\
	 				= & - \int_{0}^{2\pi} \int_{0}^{\pi}        \mathcal{M}(\theta)       \left[ \PD{\tilde{c}}{\theta}   (\IB{e}_{r} \times  \IB{e}_{\theta})  \right. \nonumber \\ & \qquad \qquad \left. +   \frac{1}{\sin \theta}  \PD{\tilde{c}}{\varphi}   (\IB{e}_{r} \times  \IB{e}_{\varphi}) \right]_{r=1}  \, \sin \theta \, \text{d}\theta \text{d}\varphi.
 				\end{align}
These chemical torques are evaluated for many configurations of JP$ _{1} $ around JP$ _{2} $ such that $\phi_2 \in [0,\pi] $ and $\phi_1 \in [-\pi,\pi]$, and consequently provides Figure \ref{fig:contour}(c,d).

Figure \ref{fig:contour}(c) shows the chemical torque contours for JP$_1$ in contact with JP$_2$ ($\epsilon=0$). The dashed region shows the configurations that were observed in the 120 experimental runs of the current study. In the contour plot, we first note that there are positive and negative peaks, representing \textit{Trans} and \textit{Cis} configurations: JP$_1$ in a \textit{Trans} orientation will experience a clockwise (positive) torque to orient its \ce{SiO2} side towards the solute abundant region, whereas during \textit{Cis} it will rotate counter-clockwise (negative). The \textit{Head-on} orientations ($\phi_1=\phi_2=0$) have zero chemical torque due to symmetry. The \textit{Ortho} orientation ($\phi_1=0, \,\phi_2=\pi/2$, as shown in figure \ref{fig:contour}(b)) experiences a CCw chemical torque. Additionally, figure \ref{fig:contour}(d) shows that increasing interparticle distance lowers the chemical torque.

We also note that the contour in figure\ref{fig:contour} (c) is not symmetric about $\phi_1=0$, it is skewed in such a manner that contacts on either immediate side of \textit{Ortho} (in the neighborhood of $\phi_2=\pi/2$) shall experience a negative (CCw) torque. This can be responsible for most \textit{Cis} and \textit{Ortho} collisions favoring \textit{Trans} departure as shown in section 3. To alternatively illustrate this skewness, figure \ref{fig:contour}(e) demonstrates the variation in chemical torque along $\phi_1$ for two configurations: $\phi_2=0$ (contact with JP$_1$ is along $-\IB{e}_z$) and $\phi_2 = \pi/4$. For the former, the chemical torque is symmetric with respect to $\phi_1=0$ (\textit{Head-on} configuration as depicted in the embedded schematic). However, the latter does not exhibit a symmetry along $\phi_1=0$ and the configuration in the embedded schematic shall experience a non-zero torque on JP$_1$, due to mobility anisotropy.  Finally, we note that out of all realized configurations, \textit{Head-on} configuration appears to be the state of lowest chemical torque. This chemical stability might be responsible for most collisions approaching each other with high $\beta$ as shown in figure \ref{fig: Contact}(c).

\section{Supporting Information}

Representative movies:\\
S1: Motion of an Isolated active JP\\
S2: JPs in \emph{O-T} collision\\
S3: JPs in \emph{H-T} collision\\
S4: JPs in \emph{T-T} collision\\
S5: JPs in \emph{C-T} collision\\
S6: JPs in \emph{T-T} collision\\
S7: JPs in \emph{non-contact} near-field interaction\\

Mean square displacement and normalized velocity autocorrelation function curves of isolated active JPs; Cumulative frequency plots ($t_{contact}$, $\beta$, $\langle|\textbf{v}_\text{inst.}|\rangle$, and $\tau_\text{r}$) and the corresponding function fit.\\

\begin{acknowledgements}
The authors acknowledge the funding received by Department of Science and Technology (SR/FST/ETII-055/2013) and the Science and Engineering Research Board (Grant numbers SB/S2/RJN-105/2017 and ECR/2018/000401), India. The authors also thank Dr. Harshwardhan Katkar from Indian Institute of Technology Kanpur for his helpful discussions.
\end{acknowledgements}

\bibliography{References}

\end{document}